\documentclass[11pt,a4paper]{article}
\usepackage{jheppub}

\usepackage{amsmath,amssymb,bm,bbm}
\usepackage{graphicx, graphics, color}
\usepackage[colorlinks=true,linkcolor=blue,citecolor=blue,urlcolor=blue]{hyperref}
\usepackage{dsfont}
\usepackage[normalem]{ulem}
\newcommand{\fref}[1]{Figure~\ref{f.#1}}

\newcommand{\eref}[1]{Eq.~(\ref{e.#1})}

\newcommand{\sref}[1]{Section~\ref{sec:#1}}

\newcommand{\beq}{\begin{eqnarray}}
\newcommand{\eeq}{\end{eqnarray}}
\newcommand{\beqs}{\begin{eqnarray*}}
\newcommand{\eeqs}{\end{eqnarray*}}

\newcommand{\diff}[1]{\mathrm{d}#1}

\newcommand{\initq}{\ensuremath{k_{i}}}

\newcommand{\xmom}{\ensuremath{k_{X}}}
\newcommand{\finalq}{\ensuremath{k_{f}}}

\newcommand{\ratcur}{\ensuremath{R_1}}
\newcommand{\rattar}{\ensuremath{R'_1}}

\newcommand{\ratgen}{\ensuremath{R_0}}
\newcommand{\ratTM}{\ensuremath{R_2}}
\newcommand{\ratX}{\ensuremath{R_3}}

\newcommand{\Tscsq}[2]{#1^2_{#2{T}}}

\newcommand{\hadpsc}{\ensuremath{P_{h}}}
\newcommand{\hadmomplain}{\ensuremath{P_{ h}}}
\newcommand{\xb}{x_{\rm Bj}}
\newcommand{\zexp}{\ensuremath{\hat{z}_N}}

\newcommand{\zex}{\ensuremath{z_N}}

\preprint{JLAB-THY-22-3550}

\title{New tool for kinematic regime estimation in semi-inclusive deep-inelastic scattering}

\author[a]{M.~Boglione,}
\affiliation[a]{Dipartimento di Fisica, Universit\`a di Torino, INFN-Sezione Torino, Italy}
\emailAdd{elena.boglione@to.infn.it}
\author[b]{M.~Diefenthaler,}
\emailAdd{mdiefent@jlab.org}
\affiliation[b]{Jefferson Lab, Newport News, Virginia 23606, USA}
\author[c]{S.~Dolan,}
\emailAdd{skd140@psu.edu}
\affiliation[c]{\mbox{Science Division, Penn State University Berks, Reading, Pennsylvania 19610, USA}}
\author[c]{L.~Gamberg,}
\emailAdd{lpg10@psu.edu}
\author[b]{W.~Melnitchouk,}
\emailAdd{wmelnitc@jlab.org}
\author[d]{D.~Pitonyak,}
\emailAdd{pitonyak@lvc.edu}
\affiliation[d]{\mbox{Department of Physics, Lebanon Valley College, Annville, Pennsylvania 17003, USA}}
\author[c,b,1]{A.~Prokudin,
\note[1]{On leave at Temple University, Philadelphia, USA}}
\emailAdd{prokudin@jlab.org}
\author[b]{N.~Sato,}
\emailAdd{nsato@jlab.org}
\author[e]{Z.~Scalyer}
\emailAdd{scalyer@p4flight.com}
\affiliation[e]{\mbox{Prepared for  Flight, LLC, York, Pennsylvania 17402, USA} \\ 
    }
\collaboration{Jefferson Lab Angular Momentum (JAM) Collaboration}

\abstract{
We introduce a new phenomenological tool based on momentum region indicators to guide the analysis and interpretation of semi-inclusive deep-inelastic scattering measurements. 
The new tool, referred to as ``affinity'', is devised to help visualize and quantify the proximity of any experimental kinematic bin to a particular hadron production region, such as that associated with transverse momentum dependent factorization. 
We apply the affinity estimator to existing HERMES and COMPASS data and expected data from Jefferson Lab and the future Electron-Ion Collider.
We also provide an interactive notebook based on Machine Learning for fast evaluation of  affinity. 
}
\begin{document}
\maketitle
\flushbottom

\section{Introduction}
\label{s.intro}

Providing a precise partonic description of hadronic structure from quantum chromodynamics (QCD) factorization theorems has been a topic of great interest for over half a century.
From inclusive and semi-inclusive deep-inelastic scattering experiments we know that hadrons have a complex internal structure involving quarks, antiquarks and gluons (generically partons) and their interactions. In addition to the partons' collinear momentum, which is highly correlated with the direction of a fast-moving parent hadron, partons also possess intrinsic transverse motion and structure.
Several types of high-energy scattering measurements are known to be sensitive to this intrinsic transverse momentum, including semi-inclusive deep-inelastic leptoproduction of hadrons $h$~\cite{Kotzinian:1994dv,Mulders:1995dh,Bacchetta:2006tn},
    \mbox{$\ell\, N \to \ell'\, h\,X$},
inclusive electron-positron annihilation to almost back-to-back hadrons~\cite{Boer:1997mf,Boer:2001he,Boer:2008fr},
    $e^+ e^- \to h_1\, h_2\, X$,
and Drell-Yan lepton-pair or weak gauge boson production in $NN$ scattering~\cite{Tangerman:1994eh},
    $N\, N\to \{\ell^+ \ell^-, Z, W^\pm\}\,X$,
where $N$ represents a proton or neutron (deuteron) in the initial state.

Interpreting these  measurements in terms of QCD requires factorization theorems that are valid for the process and the kinematic reach of the measurement.
For transverse momentum dependent (TMD) scattering reactions, TMD factorization~\cite{Collins:1981uw, Collins:1984kg,Ji:2004wu, Collins:2011zzd} describes these processes in terms of a collinear perturbative (hard) scattering cross section and nonperturbative TMD parton distribution functions (TMD~PDFs) and fragmentation functions (TMD~FFs) (collectively referred to as ``TMDs'')~\cite{Kotzinian:1994dv, Mulders:1995dh, Boer:1997nt}.
A condition implicit in the proof of TMD factorization in semi-inclusive deep-inelastic scattering (SIDIS), where at leading order the final state hadrons are fragments of the {\rm struck} quark, is that a clear separation exists between the momentum of the struck quark in the target nucleon and that of partons that are spectators to the hard collision.
In this framework the fragmentation of a quark into hadrons is independent of the production mechanism of the quark~\cite{Berger:1987zu}. 
Fragmentation is thus described by a function of the momentum fraction of the quark carried by the produced hadron, which is independent of the momentum fraction of the parent nucleon carried by the struck quark. In this scenario the hadron is said to be in the {\it current} fragmentation region.

By contrast, if the produced hadron moves in nearly the same direction as the target, the hadron is said to be in the {\it target} fragmentation region, and the relevant factorization theorem is then formulated in terms of fracture functions~\cite{Trentadue:1993ka, Grazzini:1997ih, Anselmino:2011ss,Chai:2019ykk}.
A clear distinction between the current and target fragmentation regions requires a sufficiently large separation in the momentum of the current and target fragments, and for this purpose it is convenient to use {\em rapidity} to delineate these regions. Berger~\cite{Berger:1987zu, Mulders:2000jt} provided a specific rapidity gap criterion to study the dynamics of quark fragmentation in the current fragmentation region, although in practice the delineation into distinct current, target, and  central fragmentation regions is rarely sharp~\cite{Berger:1987zu, Mulders:2000jt, Joosten:2013mia, Boglione:2016bph, Collins:2018teg}.

In addition, partons that populate the rapidity gap between current and target regions also fragment into hadrons, and these form the {\it central} fragmentation region~\cite{Collins:2018teg}.
This region can be referred to as a {\it soft-central} region, where soft gluons emitted in the cascade  after the hard scattering give important contribution to centrally produced hadrons. By soft,  we mean  partons with all four components of the momentum being of order $\mathcal{O}(m)$, where $m$ is a typical hadron mass.

Following a careful examination of the approximations involved in QCD factorization~\cite{Collins:2011zzd}, recently Boglione {\it et al.} \cite{Boglione:2016bph, Boglione:2019nwk} introduced new quantitative criteria for classifying fragmentation regions in terms of various ratios,  $R_i$,  of partonic and hadronic momenta, which are particularly useful at small and moderate values of the momentum transfer, $Q$.
Traditionally, the applicability of TMD factorization in the current region has been linked solely to the small size of the transverse momentum of the produced hadron $P_{hT}$ and the rapidity region.
It was found~\cite{Boglione:2016bph, Boglione:2019nwk}, however, that the applicability can also depend on so-called {\it region indicators}, characterized by the ratios $R_i$ 
that reflect the proximity of any given kinematic configuration to a particular partonic region of SIDIS.

Typically, in TMD phenomenology, data are filtered by the value of the hadron transverse momentum $P_{hT}$ in the Breit frame~\cite{Anselmino:2013lza, Bacchetta:2017gcc}, or by the photon transverse momentum in the hadron-hadron frame, $q_T \simeq P_{hT}/z_h$~\cite{Scimemi:2019cmh, Bacchetta:2019sam}, where
    $z_h = P\cdot P_h / P \cdot q$,
with $P$ the momentum of the initial hadron and $q$ is the momentum transfer from the incident lepton.
It was found~\cite{Boglione:2016bph, Boglione:2019nwk}, however, that cuts on $P_{hT}$ or $q_T$ applied in analyses of SIDIS data may not be sufficient to guarantee that the data, at given kinematics, are uniquely within the current fragmentation region.  
Since the observed hadrons can be produced via different physical mechanisms, identifying SIDIS cross sections in a  kinematic region corresponding to TMD factorization requires particular attention. 
It is crucial, therefore, to analyze the role that data cuts play in discriminating the current region from the target and central fragmentation regions, and assess their impact on the extraction of TMDs from future SIDIS data from Jefferson Lab (JLab), COMPASS at CERN, and the future Electron-Ion Collider (EIC). 
Indeed, application of the region indicators was already recently discussed by the HERMES Collaboration~\cite{Airapetian:2020zzo}.

In this paper we implement the region indicators introduced in Refs.~\cite{Boglione:2016bph, Boglione:2019nwk} to quantify the confidence of the proximity of SIDIS observables to a particular physical mechanism.
The new tool, which we refer to as ``affinity'', ${\cal A}$, combines information from a variety of partonic configurations and the resulting ratios, $R_i$, into a single estimate of the proximity to a particular hadron production mechanism, which ranges from 0\% to 100\%. 
We carry out the affinity analysis for kinematics relevant to existing and future facilities, and provide an affinity profile across the phase space for each kind of physical mechanism for hadron production in SIDIS~\cite{Collins:2016hqq}. 
Ultimately, these results will provide a well-defined methodology for determining the degree of confidence that a given kinematical configuration may be described in terms of TMDs, given assumptions about the partonic kinematics.

We begin in~\sref{ratios}
by briefly recalling the region indicators. To assess the proximity of the data at given kinematics to a specific physical mechanism, in~\sref{affinity} we introduce the affinity, ${\cal A}$, as a global estimator.  
In~\sref{results} we apply the new affinity tool to the analysis of existing data from the HERMES and COMPASS experiments, and discuss 
the analysis of data expected from Jefferson Lab and the future EIC.
In~\sref{interactive} we present the results of training a neural network using the {\tt TensorFlow} package, with various choices for the underlying demarcation of regions, and introduce a {\tt Google~Colab} interactive notebook for visualizing the affinity at EIC kinematics.
Finally, in~\sref{conclusions} we summarize our findings and discuss
possible future 
applications of this analysis.

\section{Region indicators}
\label{sec:ratios}

The methodology of the region indicators $R_i$ was presented in Refs.~\cite{Boglione:2016bph, Boglione:2019nwk} to help delineate different hadron production mechanisms by including some information on the underlying momentum flow in the partonic subprocess.
In this section we review the definitions of the region indicators, and discuss how they characterize the current, target and central  regions of kinematics.

\begin{figure*}[t]
\centering
\includegraphics[scale=0.6]{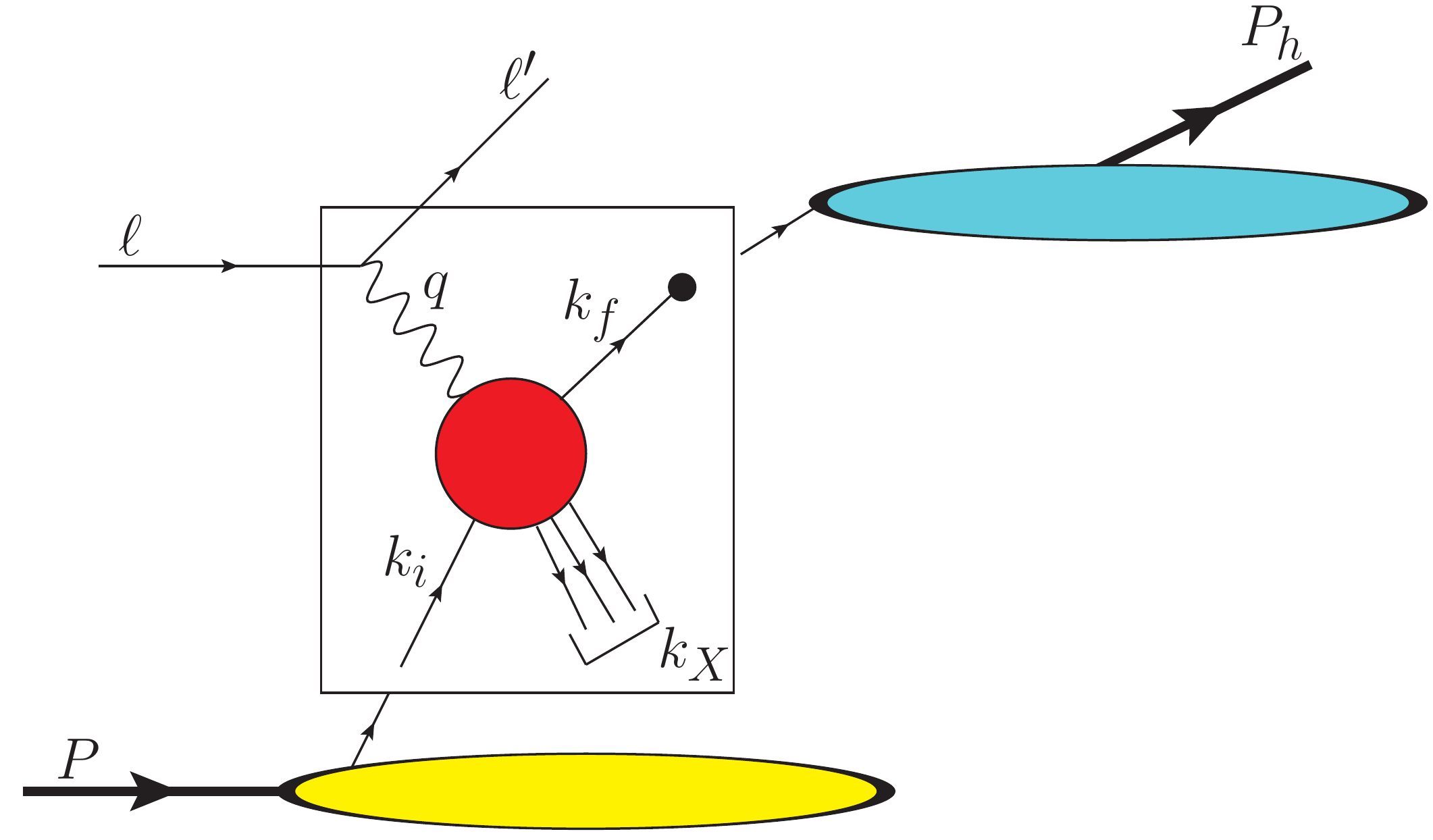}\vspace{0.5cm}

\caption{\footnotesize{Sketch of the SIDIS process, with scattering of an incident lepton (momentum $\ell$) via the exchange of a virtual photon ($q$) from a parton ($k_i$) inside a nucleon ($P$), with the scattered parton ($k_f$) fragmenting to a hadron ($P_h$) in the final state and unmeasured hadronic debris ($k_X$).
The lower (yellow) blob represents the residual system after removal of the parton from the incoming nucleon, while the upper right (cyan) blob represents the fragmentation of the outgoing parton into the observed hadron.
The rectangle envelopes the parton level subprocess, and the arrows represent momentum flow.
The black dot indicates the parton associated with the observed hadron.}}
\label{f.partonic}
\end{figure*}

\subsection{Current, target, and central regions}
\label{sec:regions}

A typical diagram for the momentum flow in the current fragmentation region of the SIDIS process is sketched in \fref{partonic}.
The figure illustrates the scattering of an incident lepton via the exchange of a virtual photon of momentum $q$ (with $q^2 \equiv -Q^2$) from an initial parton of momentum $\initq$ in a nucleon of momentum $P$ to at least one hadronizing parton $\finalq$, with $\xmom$ the total momentum of all other unobserved partons from the partonic subprocess.
A detailed discussion of SIDIS kinematics can be found in Ref.~\cite{Boglione:2019nwk}.

The current region is associated with the fragmentation of the parton after it has absorbed the incoming virtual photon. 
The outgoing parton fragments into the detected hadron of momentum $P_h$, which moves in approximately the same direction and with similar rapidity as the fragmenting parton if the transverse momentum $P_{hT}$ is small.
Consequently, in our momentum coordinate conventions the produced hadrons are in the region of negative rapidity. 
In this case, well-established TMD factorization theorems are valid --- see Refs.~\cite{Collins:1981uk, Collins:1981va, Collins:1981uw, Collins:1984kg, Meng:1995yn, Ji:2004wu, Collins:2011zzd, Aybat:2011zv, Collins:2014jpa, Echevarria:2011epo, Echevarria:2014rua}.
Hard QCD radiation may produce a large hadronic transverse momentum $P_{hT}$ in the current region, which would shift the rapidity of the hadron towards central or positive values. 
In such cases a treatment based on collinear QCD factorization theorems~\cite{Altarelli:1984pt, Collins:1984kg, Collins:2011zzd} is appropriate.

The target region is associated with the fragmentation of spectator partons, which originate in the target nucleon $P$ but do not experience the hard collision with the virtual photon.
These partons continue to move predominantly in the direction of the parent nucleon, and the products of their hadronization are found at positive values of the rapidity.
The corresponding momentum flow picture here would be similar to that in Figure~\ref{f.partonic}, but with the produced hadron originating from the lower (yellow) blob representing the remnant of the incoming nucleon.

The central 
region 
is characterized by the production of hadrons that are neither the products of a hard scattering nor associated in any obvious way with a current quark or target remnant direction. 
These hadrons are fragmentation products of quarks and gluons that fill the central rapidity region between the struck parton and the target hadron remnants~\cite{Trentadue:1993ka, Collins:2016ztc, Collins:2018teg}. 
While the identification of current and target regions is well defined by criteria which establish a clear rapidity separation between the collinear and soft sub-graphs in large-$Q$ asymptotics of factorization~\cite{Collins:2011zzd}, in reality these rapidity gaps are filled by centrally produced hadrons~\cite{Collins:2016ztc}. 
These are the phenomena that are approximated in Monte Carlo event generators by string~\cite{Sjostrand:2006za} or cluster hadronization~\cite{Corcella:2000bw}. Indeed, proofs of factorization do not yet account for graphical structures of cluster and string hadronization~\cite{Collins:2016ztc} that characterize the production of hadrons in the central region.

On the other hand, 
the fastest moving hadrons in opposite hemispheres in string-like fragmentation~\cite{Andersson:1997xwk}, being separated by space-like distances, are in line with the independent hadronization that one obtains in proofs of large-$Q$ asympotics in TMD factorization.
Additionally, as pointed out by Collins~\cite{Collins:2016ztc}, all intermediate rapidity regions between the current and target contribute at leading power in the hard scale, and it is imperative to classify the momentum regions associated with the central region. 
Thus, pinning down the central region is considerably more complicated and 
is the subject of active research~\cite{Collins:2018teg}.

One could, for instance, employ a conservative definition of the current and target regions by selecting stringent criteria for the smallness of $P_{hT}$ (or $q_T$) or the largeness of the rapidity gap. 
This is not feasible in the central region, however, where hadronization produces approximately uniform distributions in rapidity.
To define the central region one could consider identifying the region by exclusion, such that any kinematic configuration that is not strictly in the current or target regions falls within this.
Alternatively, a more conservative approach would omit slices of the process phase space by including in the central region only those configurations that are genuinely soft, according to some specific criteria (see Ref.~\cite{Collins:2018teg} for a discussion).
In either case, it is clear that in practice the boundaries of the central region remain rather ``blurred''.

To classify the relevant kinematic regions, Refs.~\cite{Boglione:2016bph, Boglione:2019nwk} proposed several ratios of partonic and hadronic momenta, as we summarize in the following.
To begin with, in order to ensure a partonic interpretation of the process, the ratio $\ratgen$ of partonic momenta to the hard scale $Q^2$, referred to as the {\it general hardness} ratio, was introduced, 
\begin{equation}
 \ratgen \equiv 
	\text{max} 
	\left( 
	\left| \frac{\initq^2}{Q^2} \right|, 
	\left| \frac{\finalq^2}{Q^2} \right|, 
	\left| \frac{\delta \Tscsq{k}{}}{Q^2} \right|
	\right)\, .
\label{e.R0}
\end{equation}
Here, $\delta \Tscsq{k}{}$ is a parameter that characterizes the size of the intrinsic transverse momentum of the parton, which is $\mathcal{O}(m^2)$, where $m$ is a typical hadron mass scale of the reaction.  
The smallness of $\ratgen$, $\ratgen \ll 1$, is the minimal requirement needed for the application of a partonic description of the SIDIS process~\cite{Boglione:2016bph, Boglione:2019nwk}.

To isolate the current fragmentation region from the target and central fragmentation regions, we define the {\it collinearity} ratio~\cite{Boglione:2016bph}, $\ratcur$, by
\begin{equation}
\ratcur \equiv \frac{\hadpsc \cdot \finalq}{\hadpsc \cdot \initq}\, .
\label{e.R1}
\end{equation}
The collinearity must be small for current fragmentation and large for target and central fragmentation.
To further distinguish the target region, we also consider the {\em target proximity} ratio $\rattar$,
\begin{equation}
\rattar \equiv \frac{\hadmomplain{}{} \cdot P}{Q^2},
\label{e.R1p}
\end{equation}
which is expected to be small for target fragmentation \cite{Boglione:2016bph, Boglione:2019nwk}.

\subsection{TMD and collinear current regions}

Historically, most phenomenological studies of SIDIS have focused on the current fragmentation region.
The analysis of this region can be refined by introducing additional ratios to distinguish the ranges of applicability of descriptions based on QCD collinear and TMD factorization theorems~\cite{Collins:2016hqq}.

For this purpose it is useful to introduce the {\it transverse hardness} ratio, $R_2$, defined as~\cite{Boglione:2019nwk}
\begin{equation}
\ratTM \equiv \frac{|k^2|}{Q^2},
\label{e.ratTM}
\end{equation}
where $k \equiv k_f-q$. 
This ratio is relevant because the $2 \to 1$ scattering process $\gamma^* q \to q'$ dominates in the TMD regime, which applies if ${|k^2|}/{Q^2} \simeq 0$.
Moreover, as shown in Ref.~\cite{Boglione:2019nwk}, one can write this ratio as
\begin{equation}
\ratTM \approx (1-\zexp)  + \zexp \frac{\Tscsq{q}{}}{Q^2}\, , 
\label{e.R2}
\end{equation}
in terms of the partonic variable $\zexp$, defined as the ratio of the ``$-$'' light-front momentum components of $k_f$ and $q$ in the Breit frame,
\begin{equation}
\zexp \equiv \frac{k_f^-}{q^-} = \frac{\zex}{\zeta}\, .
\end{equation}
The hadronic fragmentation variable $\zex$ here is defined as 
    $z_N = P_h^-/q^-$, 
with $\zeta = P_h^-/k_f^-$ the momentum fraction of the parton carried by the produced hadron in the Breit frame (see Ref.~\cite{Boglione:2019nwk} for further details).
The smallness of $\ratTM$ is needed to establish the existence of the TMD current fragmentation region.
Note that if $\Tscsq{q}{}/Q^2 \sim 1$, then $\ratTM\sim 1$ for both large and small values of $\hat{z}_N$, while if $\Tscsq{q}{}/Q^2\ll 1$ and $\zeta \sim \zex$, as in the TMD current fragmentation region, then the transverse hardness ratio becomes $\ratTM \ll 1$. On the other hand, a large value for the transverse hardness ratio $\ratTM$ would generally indicate the dominance of QCD subprocesses, such as gluon radiation, $\gamma^* q \to g q'$, to generate the observed transverse momentum $P_{hT}$. 
In Refs.~\cite{Scimemi:2019cmh, Bacchetta:2019sam} the ratio $\Tscsq{q}{}/Q^2$ was used to filter data appropriate for a TMD factorization description. 
Following Ref.~\cite{Scimemi:2019cmh}, which performs an N$^3$LO simultaneous fit of Drell-Yan and SIDIS data, in the present analysis we use the cuts
\begin{align}
Q > 2 \;{\rm GeV},\quad \frac{q_T}{Q} < 0.25
\label{eq:cutsVladimirov}
\end{align}
to select the data.
%

\begin{table}[t]
\centering
\begin{tabular}{l|c}
\hline
\hspace*{2cm} Ratio 
& Definition   
\\
\hline
~~$\ratgen$~~ general hardness  
& $\text{max} 
	\bigg( 
	\left| \dfrac{\initq^2}{Q^2} \right|, 
	\left| \dfrac{\finalq^2}{Q^2} \right|, 
	\left| \dfrac{\delta \Tscsq{k}{}}{Q^2} \right|
	\bigg)\,$      
\\ & \\
~~$\ratcur$~~ collinearity
&  $\dfrac{\hadpsc \cdot \finalq}{\hadpsc \cdot \initq}$   
\\ & \\
~~$\rattar$~~ target  proximity
& $\dfrac{\hadmomplain{}{} \cdot P}{Q^2}$
\\ & \\
~~$\ratTM$~~ transverse hardness 
& $\dfrac{\big|k^2\big|}{Q^2}$
\\ & \\
~~$\ratX$~~ spectator virtuality 
& $\dfrac{\big|\xmom^2\big|}{Q^2}$
\\ & \\
~~$R_4$~~ large transverse momentum~~~
& ~~$\text{max} 
    \bigg( 
    \left| \dfrac{\initq^2}{k^2} \right|,
    \left| \dfrac{\finalq^2}{k^2} \right|, 
	\left| \dfrac{\delta \Tscsq{k}{}}{k^2} \right|,
	\left| \dfrac{k_{iT}^2}{k^2} \right|
	\bigg)$~~
\\
\hline
\end{tabular}
\caption{\footnotesize{
Summary of the diagnostic ratios and their definitions used for identifying different fragmentation regions in SIDIS.
The particle momenta are defined as in~\fref{partonic}.}}
\label{table:ratios}
\end{table}

The region of large transverse momentum is characterized by a ratio similar to those above.
In analogy with the general hardness ratio $R_0$ in Eq.~(\ref{e.R0}), we introduce the {\it large transverse momentum} ratio, $R_4$,
\begin{subequations}
\begin{align}
R_4 & \equiv 
	\text{max} \left( \left| \frac{\initq^2}{k^2} \right|, \left| \frac{\finalq^2}{k^2} \right|, 
	\left| \frac{\delta \Tscsq{k}{}}{k^2} \right|, \left|\frac{k_{iT}^2}{k^2} \right|\right) 
\\
	& = \frac{1}{R_2}  \text{max} \left( \left| \frac{\initq^2}{Q^2} \right|, \left| \frac{\finalq^2}{Q^2} \right|, 
	\left| \frac{\delta \Tscsq{k}{}}{Q^2} \right|, \left|\frac{k_{iT}^2}{Q^2} \right|\right),
\label{e.R4}
\end{align}
\end{subequations}
using the definition of $R_2$ in Eq.~(\ref{e.ratTM}).
Transverse momentum can be said to be generated by perturbative mechanisms if $R_4 \ll 1$. 
The smallness of $R_4$ will be used in this analysis to determine the extent of the collinear QCD current region, instead of the requirement that the transverse hardness ratio $\ratTM$ be large.

We can also explore the region associated with gluon radiation in more detail by introducing the {\em spectator virtuality} ratio, $\ratX$, defined by
\begin{equation}
\ratX \equiv  \frac{|\xmom^2|}{Q^2} \, ,
\label{e.R3}
\end{equation}
where $k_X = k_i+q-k_f$.
Small values of $\ratX$ correspond to $2 \to 2$ parton kinematics, and the corresponding region may be explained by low-order (LO) perturbative QCD (pQCD) calculations.
On the other hand, large $\ratTM$ and $\ratX$ values correspond to $2 \to n$ scattering, where $n \ge 3$, so that higher-order (HO) pQCD calculations are needed to describe data in this region.

Finally, the region of matching of TMD and collinear factorizations is characterized by the presence of intermediate values of $\ratTM$, so that both the TMD and collinear pictures are approximately valid, and a smooth transition between these is possible.
For completeness, in Table~\ref{table:ratios} we summarize the definitions of all the ratios that act as region indicators in SIDIS analysis.

In addition to the transverse hardness ratio, it is also useful to consider the logarithm measure, $|\ln R_2|$, which is typical of the type of large logarithm that requires the $q_T$-resummation component from the Collins-Soper-Sterman (CSS) treatment of evolution when $R_2 \to 0$.
If the logarithm measure $|\ln R_2|$ becomes larger than ${\cal O}(1)$, then $q_T$-resummation effects may need to be taken into account.

\begin{table}[t]
\begin{tabular}{l|c|c|c|c|c|c}
\hline
~Region~ &~ $\ratgen$ & $\ratcur$ & $\rattar$ & $\ratTM$  & $\ratX$ & $R_4$  
\\
\hline
~TMD
& ~small~   & small     & $\times$  & ~small~   & $\times$  & $\times$ 
\\
~matching~ &
~small~     & small     & $\times$  & small     & $\times$  & $\times$  
\\

~collinear 
& ~small~   & small     & $\times$  & large     & ~small (LO pQCD)~ & ~small~
\\

&    &     &   &      & ~large (HO pQCD)~ & 
\\
\hline
~target 
& ~small~   & large  &  small & $\times$ & $\times$ & $\times$
\\
\hline
~central 
& ~small~   & ~not small~ & ~not small~ & small & $\times$ & $\times$  \\
\hline
\end{tabular}
\caption{\footnotesize{Examples of sizes of region indicator ratios corresponding to particular regions of SIDIS. The symbol ``$\times$'' denotes ``irrelevant or ill-defined'' (see text for further details).}}
\label{table:catalogue}
\end{table}

The resulting catalogue of possible values of region indicators is presented in Table~\ref{table:catalogue}. 
As shown, the proximity of a given physical mechanism is characterized by the different sizes of the $R_i$ ratios, which in turn depend not only on the external kinematics of the SIDIS reaction, but also on the internal active parton momenta.
Since the latter are not physical observables, the use of $R_i$ requires prior knowledge of the parton momenta, which can be inferred from nonperturbative treatments of QCD or from phenomenological analyses where the proximities of regions are estimated on the basis of agreement between data and theory.

\section{Affinity}
\label{sec:affinity}

To facilitate the assessment of the proximity of data at a given set of kinematics to a specific physical mechanism, we introduce \emph{affinity}, ${\cal A}$, as a global estimator using a Bayesian formulation,
\begin{align}
{\cal A}\big(\xb,Q^2,&z_h,P_{hT}\big|{\rm region}\big)
=\!\int\!\diff \{R_i\} \;\;\Theta\big(\{R_i\}\big|~{\rm region}\big)
\notag\\
&\times 
\!\!\int \diff^4 k_i\, \diff^4 k_f\, \diff^4 \delta k_{\rm T} 
 ~{\cal P}\big(\{R_i\}\big|\xb,Q^2,z_h,P_{hT};k_i,k_f,\delta k_T\big)
 ~\pi\big(k_i,k_f,\delta k_T\big),
\label{e.affinity}
\end{align}
where $\xb = {Q^2}/{2 P \cdot q}$ is the Bjorken scaling variable. 
The second line of \eref{affinity} is a joint distribution for the SIDIS $R_i$ indicators marginalized over a given choice for the prior distribution $\pi$, with $\mathcal{P}$ the conditional probability density for $\{R_i\}$. 
The latter is a function chosen according to the \emph{prior beliefs} for the possible ranges of intrinsic  
partonic momenta.
Given such a joint distribution, the affinity of a given SIDIS kinematic bin to a given region is defined by marginalizing the joint distribution over all possible values of $\{R_i\}$ with the \emph{proximity function} $\Theta$ that selects a given region in ${R_i}$-space, according to Table~\ref{table:catalogue}.

In practice, the implementation of \eref{affinity} requires the use of Monte Carlo methods, in which one must sample the four-vectors of the parton momenta. 
Using light-front coordinates, we parametrize the initial and final parton four-momenta as
    $k_i=\big( k_i^+, k_i^-, \boldsymbol{k}_{iT}\big)$ and
    $k_f=\big( k_f^+, k_f^-, \boldsymbol{k}_{fT}\big)$,
and the intrinsic transverse momentum of the parton as
    \mbox{$\delta k_T = \big( 0, 0, \delta\boldsymbol{k}_T \big)$},
with the components given by
\begin{align}
k_i^+ &= \frac{Q}{\sqrt2\, \hat{x}_N},\quad k_i^- = \frac{\hat{x}_N}{\sqrt2\, Q}
         \big(k_i^2 + \boldsymbol{k}_{iT}^2 \big),\quad
\boldsymbol{k}_{iT}
= \big( |\boldsymbol{k}_{iT}| \cos\varphi_{k_i},~
         |\boldsymbol{k}_{iT}| \sin\varphi_{k_i}
   \big),\\ \nonumber \\
k_f^+
&=\frac{1}{\sqrt2\, Q\hat{z}_N}
   \big(     k_f^2
        + (q_T\, \hat{z}_N \sin\varphi_h
            - |\delta\boldsymbol{k}_T| \sin\varphi_{\delta k})^2
        + (q_T\, \hat{z}_N \cos\varphi_h
            - |\delta\boldsymbol{k}_T| \cos\varphi_{\delta k})^2
   \big),
\nonumber \\
k_f^- &= \frac{Q}{\sqrt2\, \hat{z}_N},\quad
\boldsymbol{k}_{fT}
= \big( |\delta\boldsymbol{k}_T| \cos\varphi_{\delta k}
            - q_T\, \hat{z}_N \cos\varphi_h,~
         |\delta\boldsymbol{k}_T| \sin\varphi_{\delta k}
            - q_T\, \hat{z}_N \sin\varphi_h
   \big),
\\ \nonumber \\
\delta\boldsymbol{k}_T
&=  \big(
        |\delta\boldsymbol{k}_T| \cos\varphi_{\delta k},~
        |\delta\boldsymbol{k}_T| \sin\varphi_{\delta k}
   \big),
\end{align}
where  
$x_N\equiv -q^+/P^+=2\xb/(1+\sqrt{1+4 \xb^2M^2/Q^2})$ and $\hat{x}_N=x_N/\xi$.
In this form, the partonic momenta are parametrized in terms of the two momentum fractions $0< \xi,\zeta < 1$, two invariant masses $k_i^2 < 0$ and $k_f^2 > 0$, two transverse momenta $|\boldsymbol{k}_{iT}|$, $|\delta\boldsymbol{k}_{T}|>0$, and two partonic angular variables $\varphi_{k_i}$ and $\varphi_{\delta k}$ in the Breit frame, as well as the external kinematic variables $x_N$, $z_N$, $Q$, $q_T$, and~$\varphi_h$.
In terms of these variables, we can write \eref{affinity} as
\begin{align}
{\cal A}\big(\xb,Q^2,z_h,P_{hT} \big|\, &{\rm region}\big)
=\!\int\!\diff \{R_i\}\;
    \Theta\big(\{R_i\}\big|\, {\rm region}\big)
\notag\\
&\times 
\int\! \diff \xi ~\diff \zeta
~\diff k_i^2 ~\diff k_f^2
~\diff |\boldsymbol{k}_{iT}|
~\diff|\delta\boldsymbol{k}_T|
~\diff \varphi_{k_i}
~\diff \varphi_{\delta k}
~\diff \varphi_h
\notag\\
&\times\, 
{\cal P} \big( \{R_i\}\, \big|\, \xi,\zeta,k_i^2,k_f^2,|\boldsymbol{k}_{iT}|, |\delta\boldsymbol{k}_T|,\varphi_{k_i}, \varphi_{\delta k} \big)
\notag\\
&\times\,
\pi \big( \xi,\zeta,k_i^2,k_f^2,|\boldsymbol{k}_{iT}|, |\delta\boldsymbol{k}_T|,\varphi_{k_i}, \varphi_{\delta k}\,
    \big|\,
    \xb,Q,z_h,q_T,\varphi_h
    \big)\; .
\label{e.affinity2}
\end{align}
As given, \eref{affinity2} is quite general up to the freedom of choosing the priors and the proximity function.
The priors for the parton momenta are selected as follows:
\begin{itemize}
    \item \underline{Momentum fractions}: Since the physical values of $\xi$ and $\zeta$ are bounded in the ranges $\xb<\xi<1$ and $z_h<\zeta<1$, and the lowest order diagrams have $\xi=\xb$ and $\zeta=z_h$, we use flat priors within a window $\epsilon$ around $\xb$ and $z_h$,
    \begin{align}
    \pi(\ldots) \supset  \theta(\xb<\xi<\xb+\epsilon)
                        ~\theta(z_h<\zeta<z_h+\epsilon),
    \end{align}
    where $\theta$ is the Heaviside step function, and we take with $\epsilon=0.1$.
    
    \item \underline{Angular variables}: While angular modulations become nontrivial for spin-dependent observables, we restrict our analysis to unpolarized SIDIS reaction and therefore use flat priors,
    \begin{align}
    \pi(\ldots) \supset  \theta(0<\varphi_{k_i}<2\pi)
                        ~\theta(0<\varphi_{\delta k}<2\pi)
                        ~\theta(0<\varphi_{h}<2\pi).
    \end{align}

    \item \underline{Parton virtualities and intrinsic transverse momenta}: We use Gaussian distributions ${\cal G}$ to estimate the parton dynamics in the infrared region,
    \begin{align}
    \pi(\ldots) \supset 
    ~{\cal G}\Big( {\sqrt{|k_i^2|}}\, \Big|\, m,\Delta\Big)
    ~{\cal G}\Big( {\sqrt{|k_f^2|}}\, \Big|\, m,\Delta\Big)
    ~{\cal G}\big( |\delta \boldsymbol{k}_{T}| \big| m,\Delta \big)
    ~{\cal G}\big( |\boldsymbol{k}_{i,T}|      \big| m,\Delta),
    \end{align}
    with mean $m=0.5$~GeV and width $\Delta=0.5$~GeV guided by nonperturbative QCD fits of TMD widths~\cite{Anselmino:2013lza, Signori:2013mda}, relevant for the experiments we consider in this paper.
\end{itemize}
While more sophisticated choices for the these distributions can be made, \eref{affinity2} will be sufficient to illustrate the practical use of affinity.
For our proximity function we use a flat distribution of the form
\begin{align}
\Theta\big( \{R_i\}\, \big|\, {\rm region} \big)
= \prod_i \theta\big( R_i<R^{\rm max}_i({\rm region}) \big)\,
          \theta\big( R_i>R^{\rm min}_i({\rm region}) \big)\,
\end{align}
that isolates the desired region. 
The first $\theta$ function is used to define the region of ``small'' values and the second $\theta$ is used to define the region of ``large'' values. 
Note that both $\theta$ functions are not always needed for each region. In the case of the TMD-collinear transition region, we will also use $\theta\big( R^{\rm max}_2({\rm region}) < R_2 < R^{\rm min}_2({\rm region})\big)$  for $R_2$ to define intermediate values.

Since for reasonable values for $R^{\rm min}_i$ and $R^{\rm max}_i$ we have {\it a priori} no quantitative knowledge of specific region boundaries other than the qualitative estimates in Table~\ref{table:catalogue}, in practice we need to appeal to existing TMD phenomenology for guidance. 
Specifically, we tune the allowed ranges of $R_i$ such that for the kinematic bins where Ref.~\cite{Scimemi:2019cmh} found a good agreement between data and phenomenology, the affinity ${\cal A} \sim 1$, and for the excluded kinematic bins the affinity ${\cal A} \sim 0$.

In terms of the kinematic cuts in Eq.~\eqref{eq:cutsVladimirov}, our implementation translates as ``small'' $R_i$ values, with $R^{\rm max}_i({\rm TMD}) = 0.3$, for $i=0,1,2$.
As there are no studies of collinear, central or target regions available to quantify ``large'' values, we will define as ``large'' any value that is at least 3 times greater than ``small'', so that 
    $R^{\rm min}_i({\rm region}) = 0.9$, for $i=0,1,2$.
For other ratios we set 
    $R^{\rm max}_i({\rm region}) = 0.3$ and
    $R^{\rm min}_i({\rm region}) = 3R^{\rm max}_i({\rm region}) = 0.9$.
These values in principle may vary depending on kinematics, so that more fine tuning may be required to delineate the regions with greater accuracy.

\section{Applications}
\label{sec:results}

\begin{figure}[t] 
\centering
\includegraphics[width=0.6\textwidth]{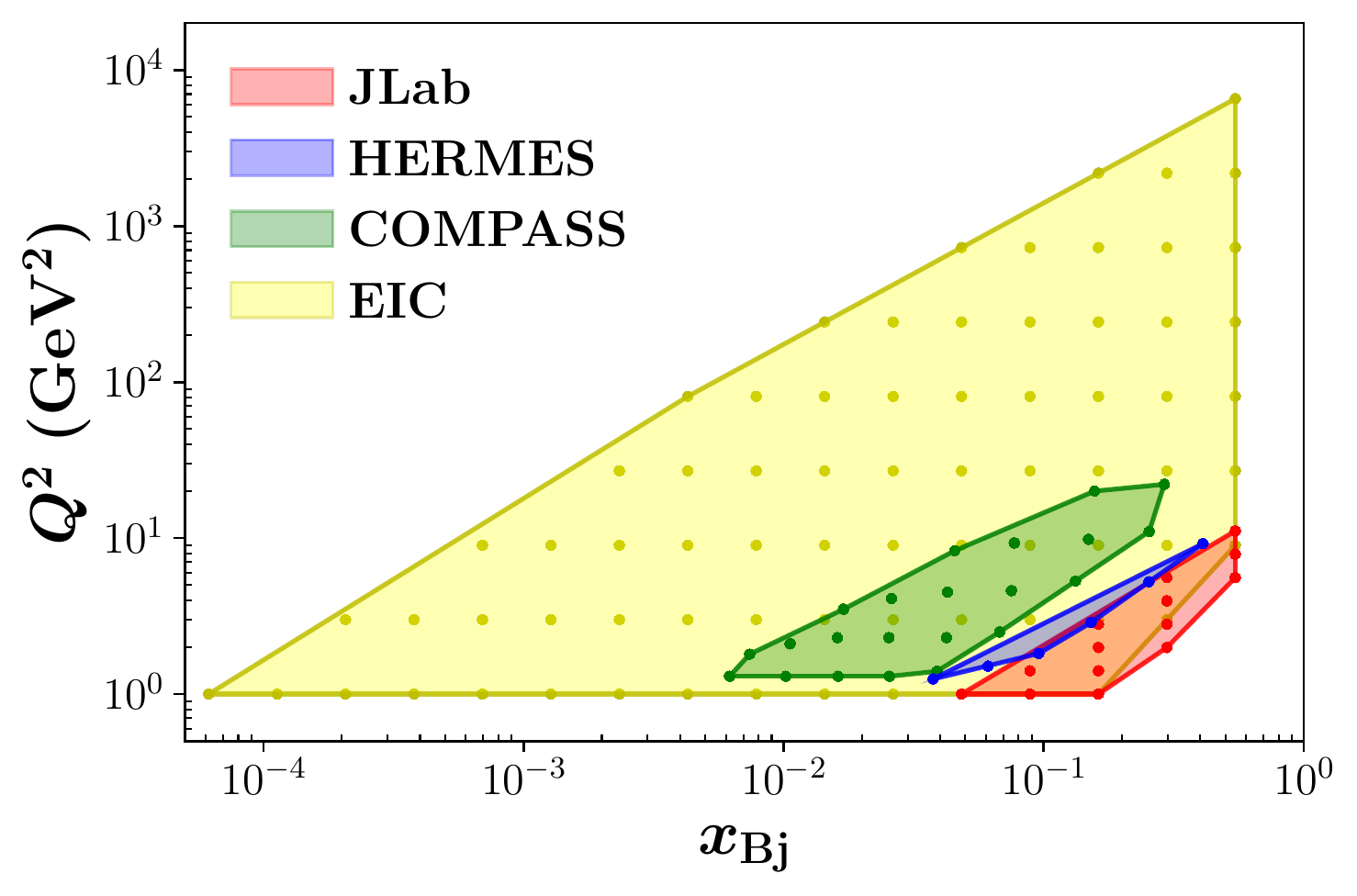}
\caption{\footnotesize{Kinematical reach of $Q^2$ (GeV$^2$) versus $\xb$ for data from existing experiments at Jefferson Lab (red), HERMES (blue), COMPASS (green), and the future EIC (yellow). Bin centers are indicated by filled circles, with each bin representing measurements for different values of $z_h$ and~$P_{hT}$.}}
\label{f.experiments_kinematics}
\end{figure}

In this section we demonstrate the practical utility of the region indicators introduced above by applying our operative definition of affinity to data from existing experiments, as well as to expected data from future facilities.
In \fref{experiments_kinematics} we illustrate the kinematic reach in $\xb$ and $Q^2$ of the experiments considered in this paper, namely, Jefferson Lab, HERMES, COMPASS, and the future EIC, with measurements to be performed at points with bin centers in $\xb$ and $Q^2$ indicated.
We will study the kinematic reach of the experiments in terms of the produced hadron rapidity, defined by
    $$y_h\equiv \frac12 \ln \left|\frac{P_h^+}{P_h^-}\right|\;.$$
While rapidity is an observable, it can be challenging to measure, particularly at large values, where particle trajectories are close to the beam pipe and neither their energies nor their total momenta can be precisely determined.
In practice, pseudorapidity is often used instead, which is a function of the polar angle between the particle trajectory and the beam axis, and thus ideal for discussions of acceptance coverage of collider detectors and the placement of their various components. 
For highly relativistic particles, rapidity and pseudorapidity are almost identical and both can be used for physics discussions.

\begin{figure}[t] 
\centering
\includegraphics[width=0.65\textwidth]{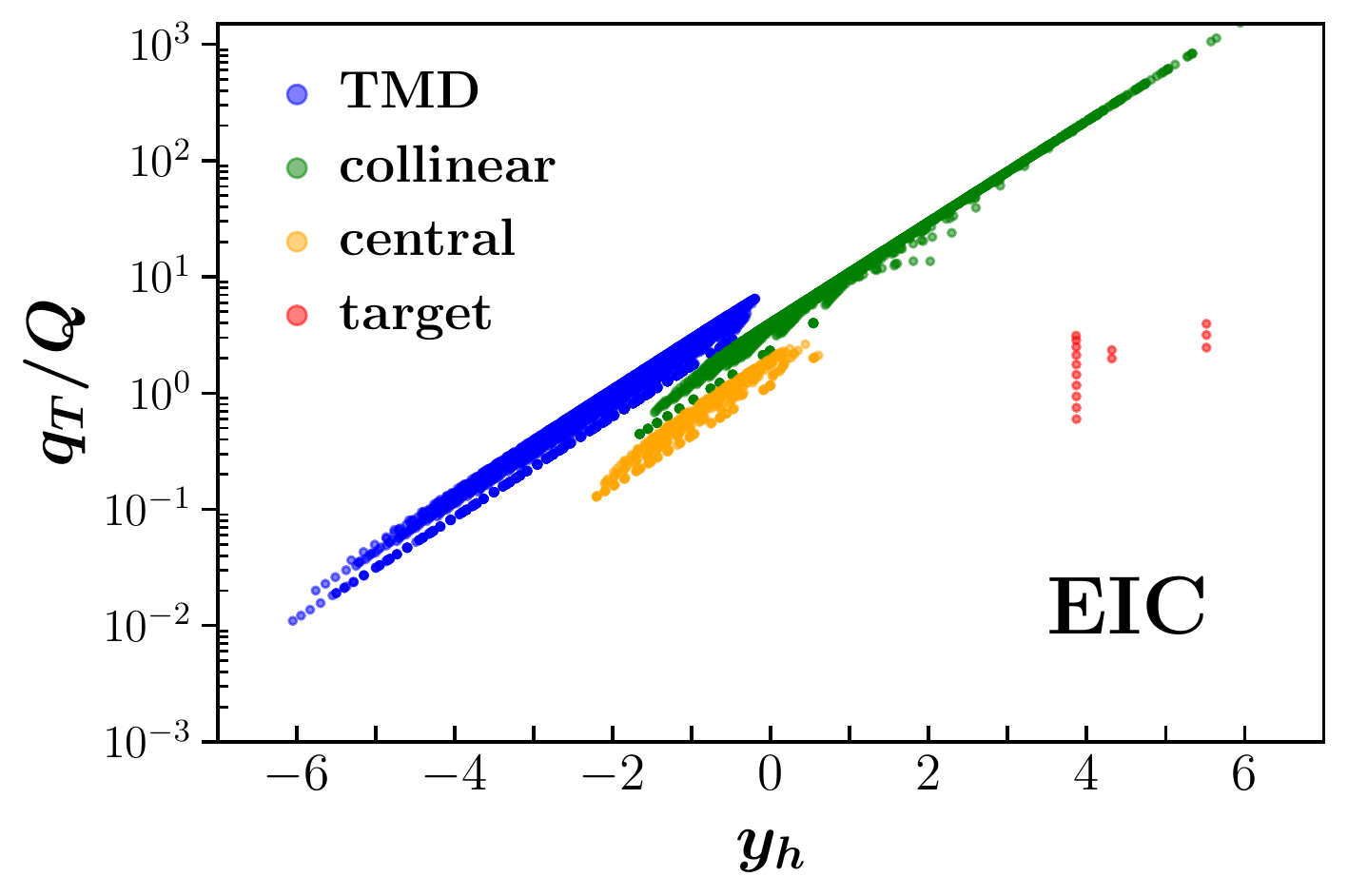}
\caption{\footnotesize{
Distribution of bins corresponding to the TMD (blue), collinear (green), central (yellow), and target (red) regions as a function of $q_T/Q$ and rapidity $y_h$. To make the points distinguishable in regions of overlap, we scale $q_T/Q$ in the TMD region by a factor 8, in the collinear region by 4, and in the central region by 2.
}}
\label{f.EIC_ratios}
\end{figure}

For typical kinematics expected at the future EIC, with variable center of mass energy from $\sqrt{s}=20$~GeV to $140$~GeV, we simulated semi-inclusive deep-inelastic $\pi^+$ production for 7400 bins in $\xb$, $z_h$, $Q^2$, and $P_{hT}$, as in the EIC Yellow Report EIC~\cite{AbdulKhalek:2021gbh}. 
\fref{EIC_ratios} shows the kinematics of the EIC projected data, categorized according to affinity values exceeding the threshold of 5\% for various fragmentation regions. 
The initial proton is always in the positive rapidity range, while the produced hadron has either positive or negative rapidity.
As discussed above, hadrons produced at negative rapidity are likely to be in the current fragmentation region.
Hadrons with higher values of $q_T$ migrate into the central and positive rapidities and may originate from hard gluon scattering; therefore, they will be described by collinear QCD. 
The central region, where low energy partons hadronize, is likely to be in the intermediate region of rapidity.
Finally, the target region is typically associated with hadrons with positive  rapidities.

As \fref{EIC_ratios} demonstrates, the region of central rapidities $y_h \sim 0$ corresponds to an admixture of almost all regions.
Although the ratio $q_T/Q$ appears to be a good indicator for separating the TMD and collinear regions, a residual overlapping among central, collinear and TMD regions can only be resolved by accounting also for the value of the hadron rapidity.
We note that there are two solutions for $y_h$ [see Eq.~(20) of Ref.~\cite{Boglione:2016bph}], which are on opposite sides of the proton rapidity, and the solution that corresponds to the target fragmentation region is severely constrained by kinematics.
The final state hadron has a mass smaller than that of the proton, and if $P_{hT}$ is small enough, then $z_h$ will be small.
One can see from \fref{EIC_ratios} that target region bins are located in the positive range of rapidity, and $q_T/Q$ is small.
We will see below that values of $z_h$ are also small for the target fragmentation region.

\begin{figure}[t] 
\centering
\includegraphics[width=0.5\textwidth]{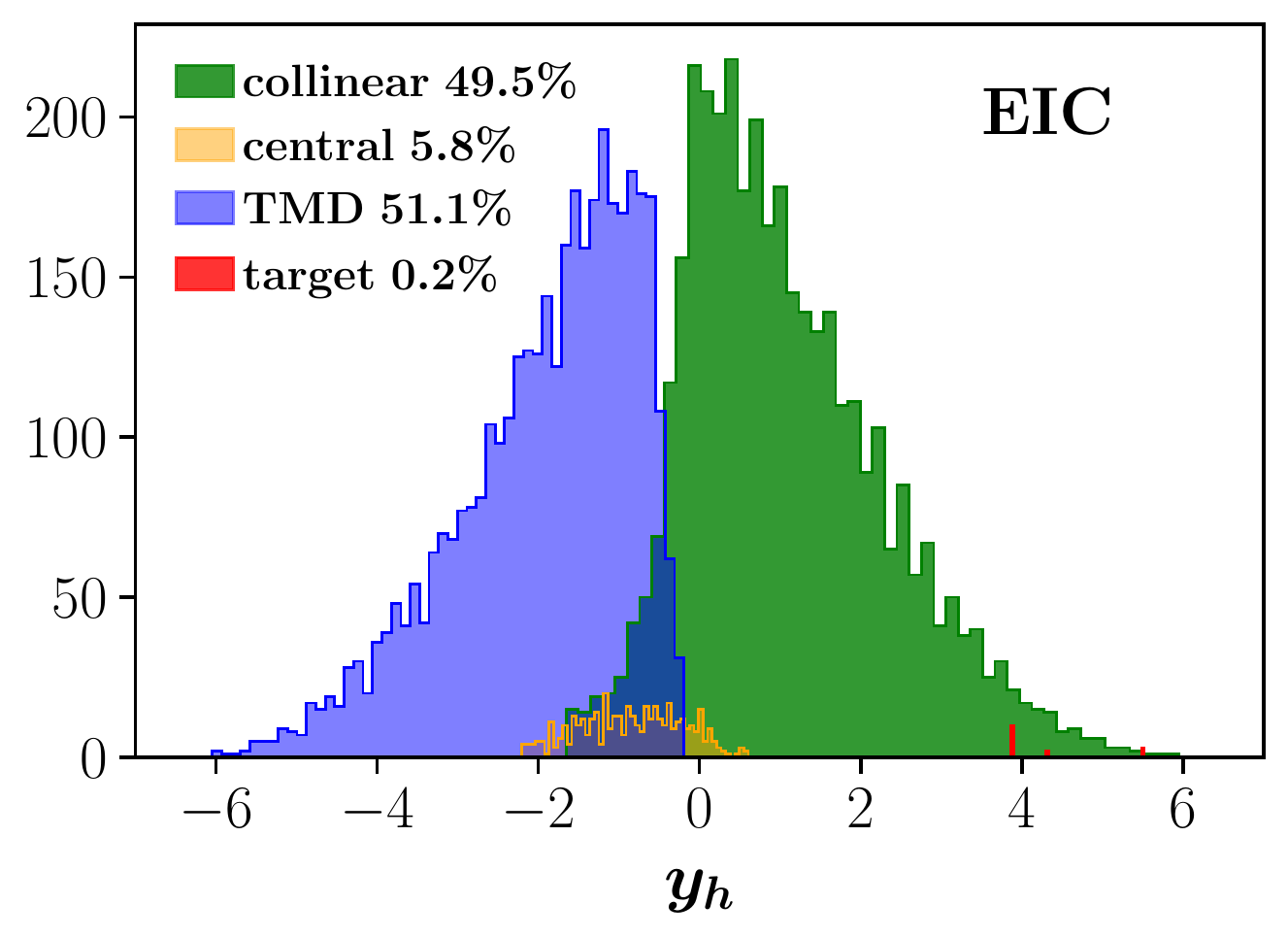}
\includegraphics[width=0.49\textwidth]{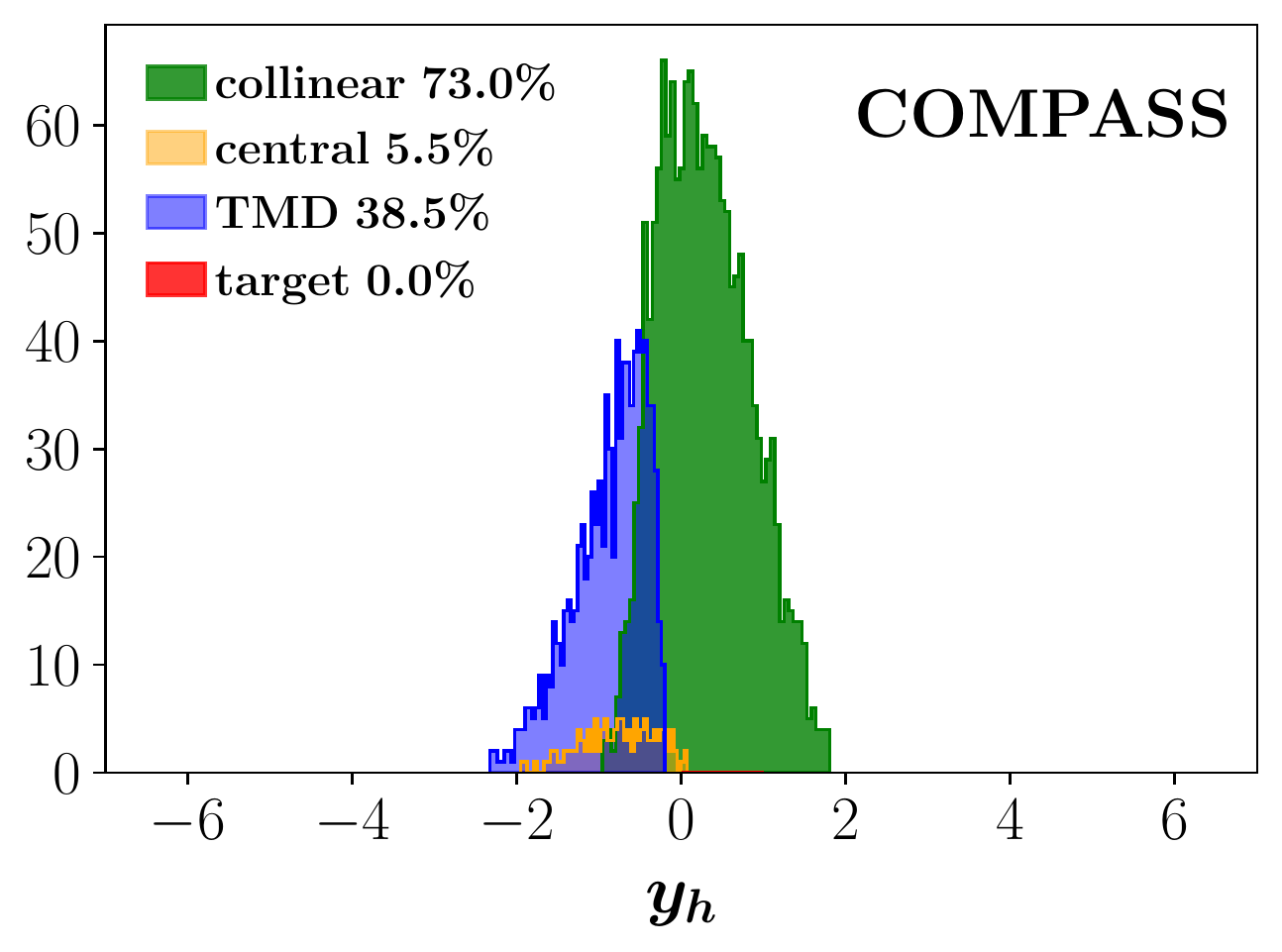}
\includegraphics[width=0.49\textwidth]{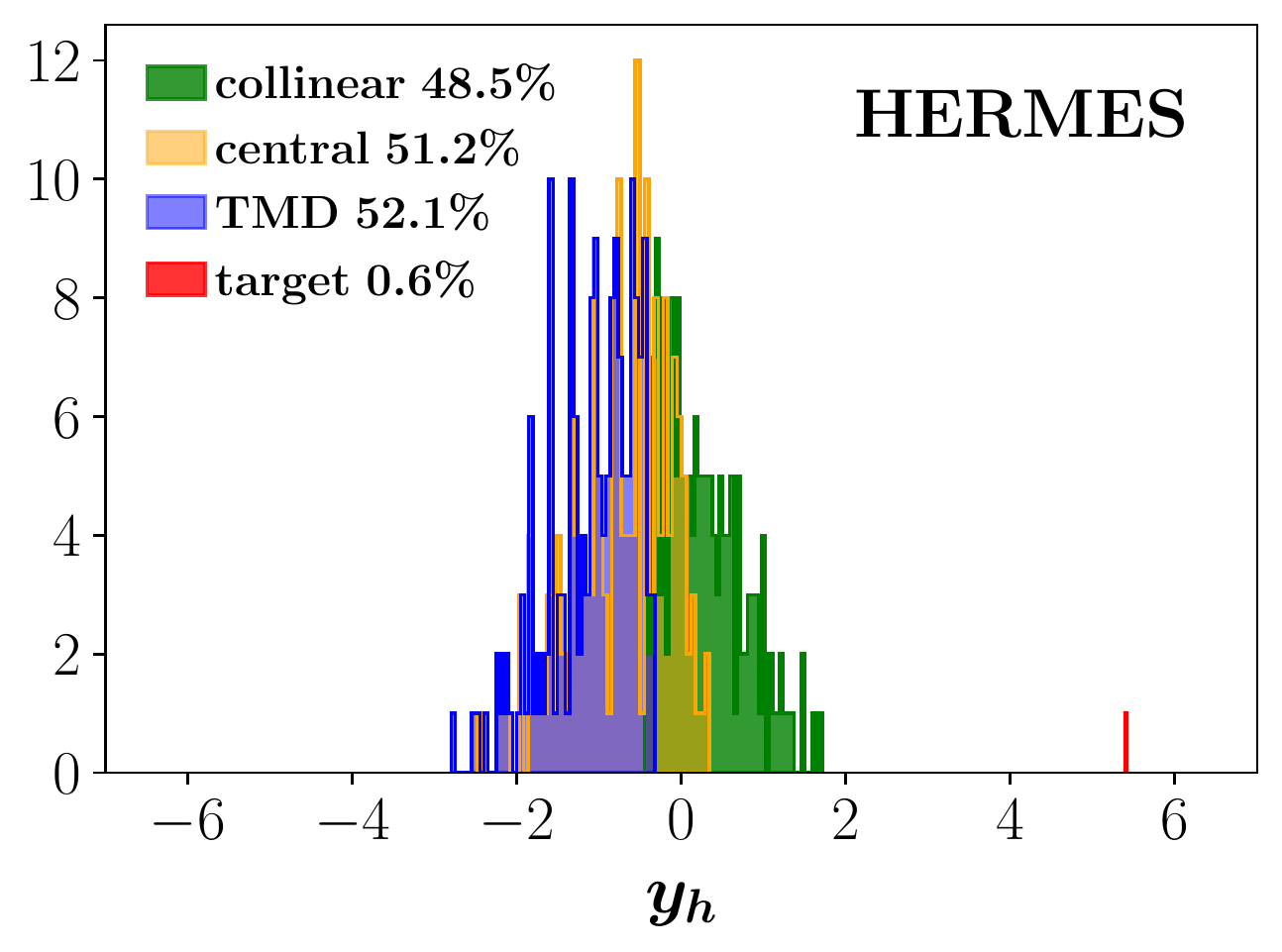}
\includegraphics[width=0.49\textwidth]{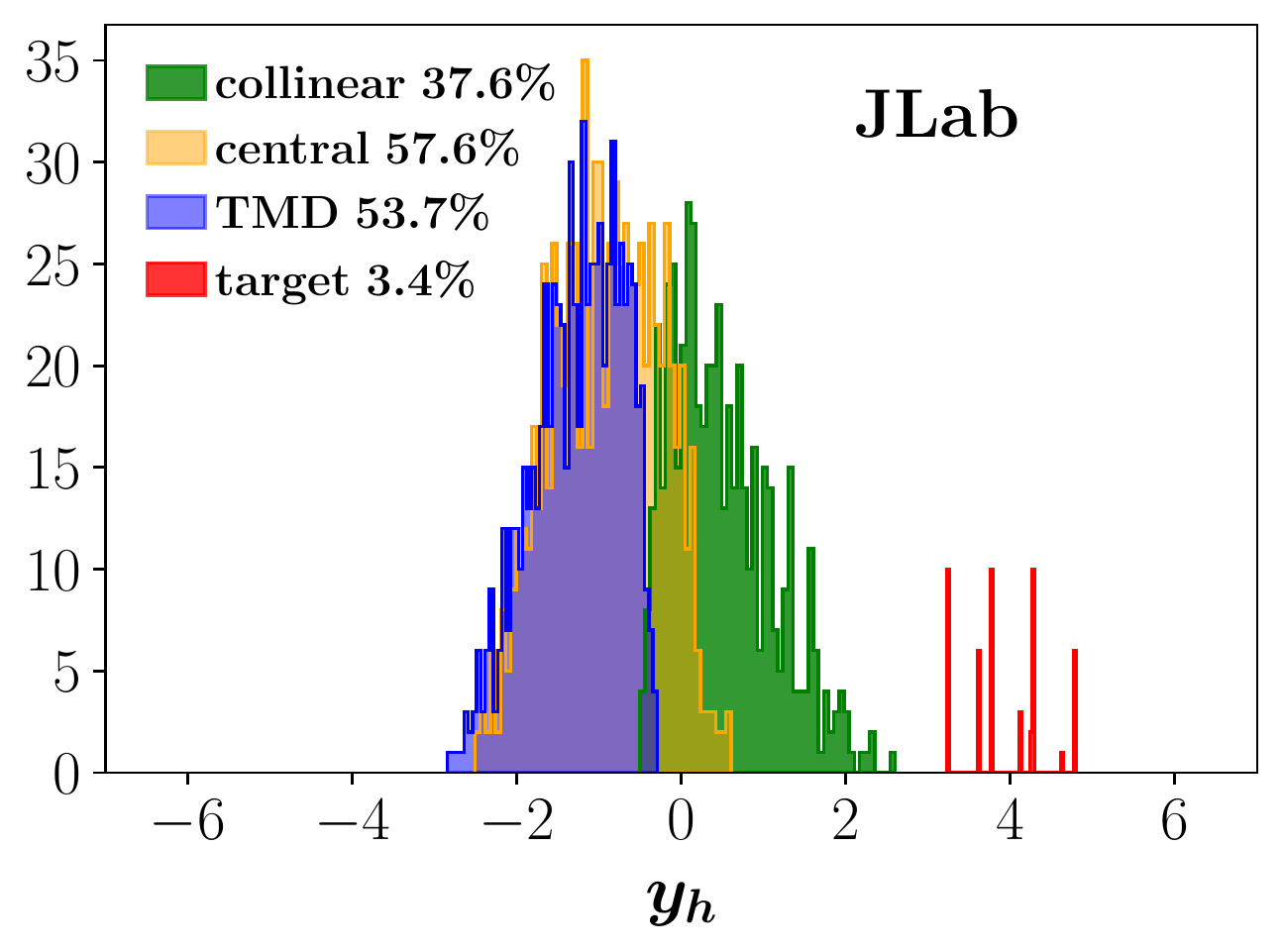}
\caption{\footnotesize{Phase space in rapidity $y_h$ of produced hadrons at EIC, COMPASS, HERMES, and Jefferson Lab kinematics, with
for TMD (blue), collinear (green), central (yellow) and target (red) regions indicated. The legends show the percentage of all bins with corresponding affinity above 5\%.}}
\label{f.exp_dist}
\end{figure}

The distribution of all kinematic bins accessible at current and future facilities is shown in \fref{exp_dist} as a function of the produced hadron rapidity, $y_h$, categorized by the affinity as in \fref{EIC_ratios}. 
One can see that at the EIC the majority of the data will correspond to either the TMD or collinear QCD fragmentation regions, with small fractions of events in the soft and target fragmentation regions.

At other, lower energy facilities, such as COMPASS, HERMES, and Jefferson Lab, the reach in rapidity is clearly smaller.
At Jefferson Lab kinematics, for example,
where the measurements are at center of mass energy $\sqrt{s} = 4.6$~GeV, one is likely to encounter larger portions of events from central and target fragmentation regions.
At the same time, one expects to have large fractions of events that correspond to TMD and collinear factorization for all the experiments discussed.
Note also that the regions can overlap; consequently, the sum of percentages for affinities does not equal 100\%.
We will study each region in more detail in the following. 

\subsection{TMD region}
\label{sec:resultstmd}

\begin{figure}[t] 
\hspace*{-0.9cm}\includegraphics[width=1.1\textwidth]{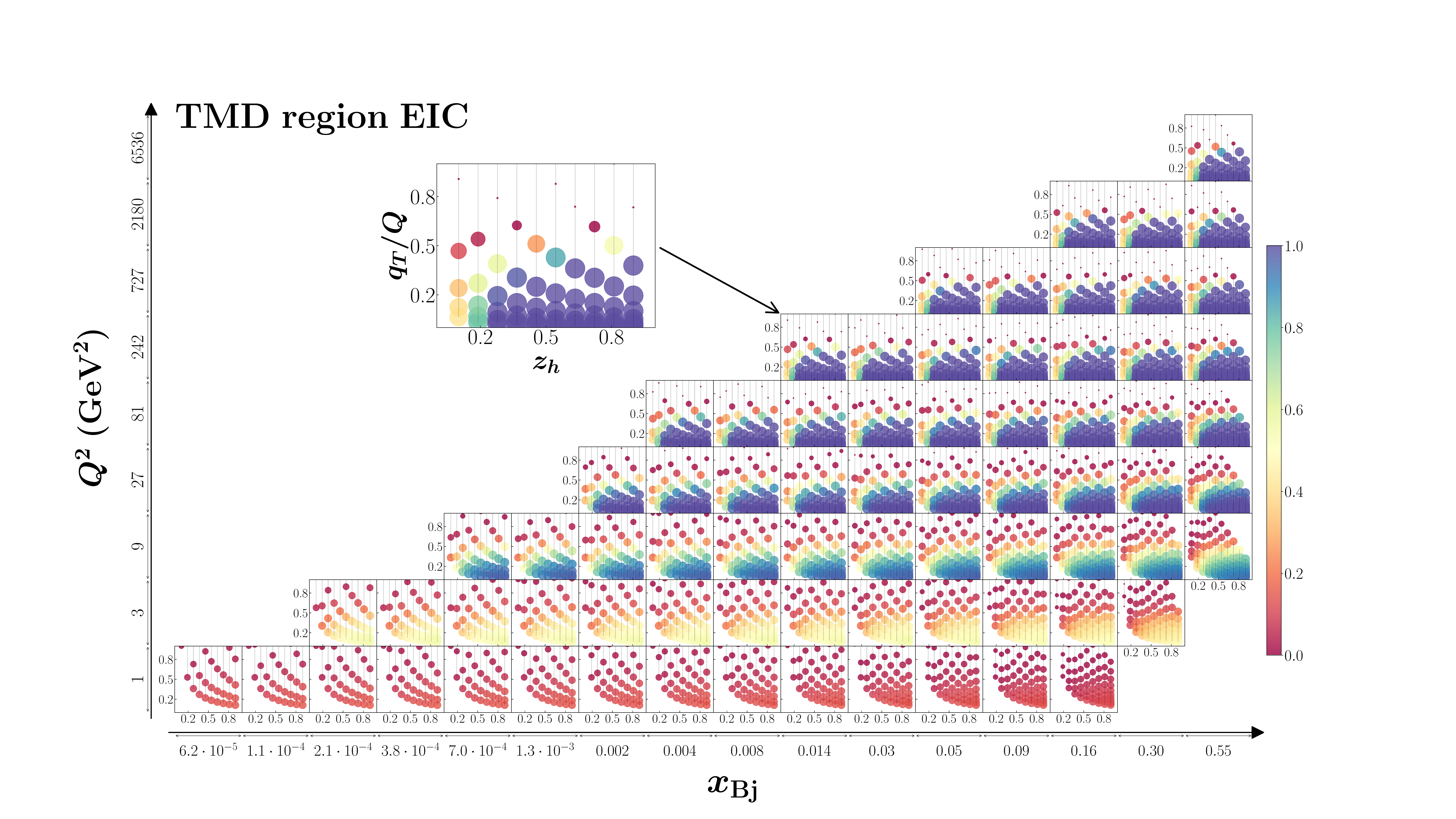}
\caption{TMD affinity for EIC kinematics. Bin centers are located in the points corresponding to the bin averaged values of $\xb$ and $Q^2$, and in each of these bins various values of $z_h$ and $q_T/Q$ can be measured. In each bin of fixed $z_h$ and $q_T/Q$, the affinity is indicated by a dot with size proportional to the corresponding affinity value. The affinity is color coded according to the scheme on the right of the panels: red (and smaller) symbols correspond to low TMD affinity, while dark blue (and larger) symbols correspond to high TMD affinity.}
\label{f.EIC_tmd}
\end{figure}

TMD affinity is calculated using Eq.~\eqref{e.affinity} by requiring the region indicators $\ratgen$, $\ratcur$, and $\ratTM$ to be small. 
The results for the bins at the EIC kinematics are shown in \fref{EIC_tmd}.
One can see that bins with relatively large $\xb$ and $Q^2$ values (and relatively high $z_h$ and~$P_{hT}$) are particularly important for the TMD factorization description.
In terms of the applicability of TMD factorization, this suggests that $q_T/Q$ becomes sufficiently small for the factorization to be valid.
We estimate 2325 out of the 7400 bins to have TMD affinity of 68\% or higher and 1739 bins to have TMD affinity of 95\% or higher.
As discussed below, the remainder of the data (or at least part of it) will correspond to different mechanisms, such as those associated with the collinear factorization scheme.

\begin{figure}[t!] 
\centering
\includegraphics[width=0.75\textwidth]{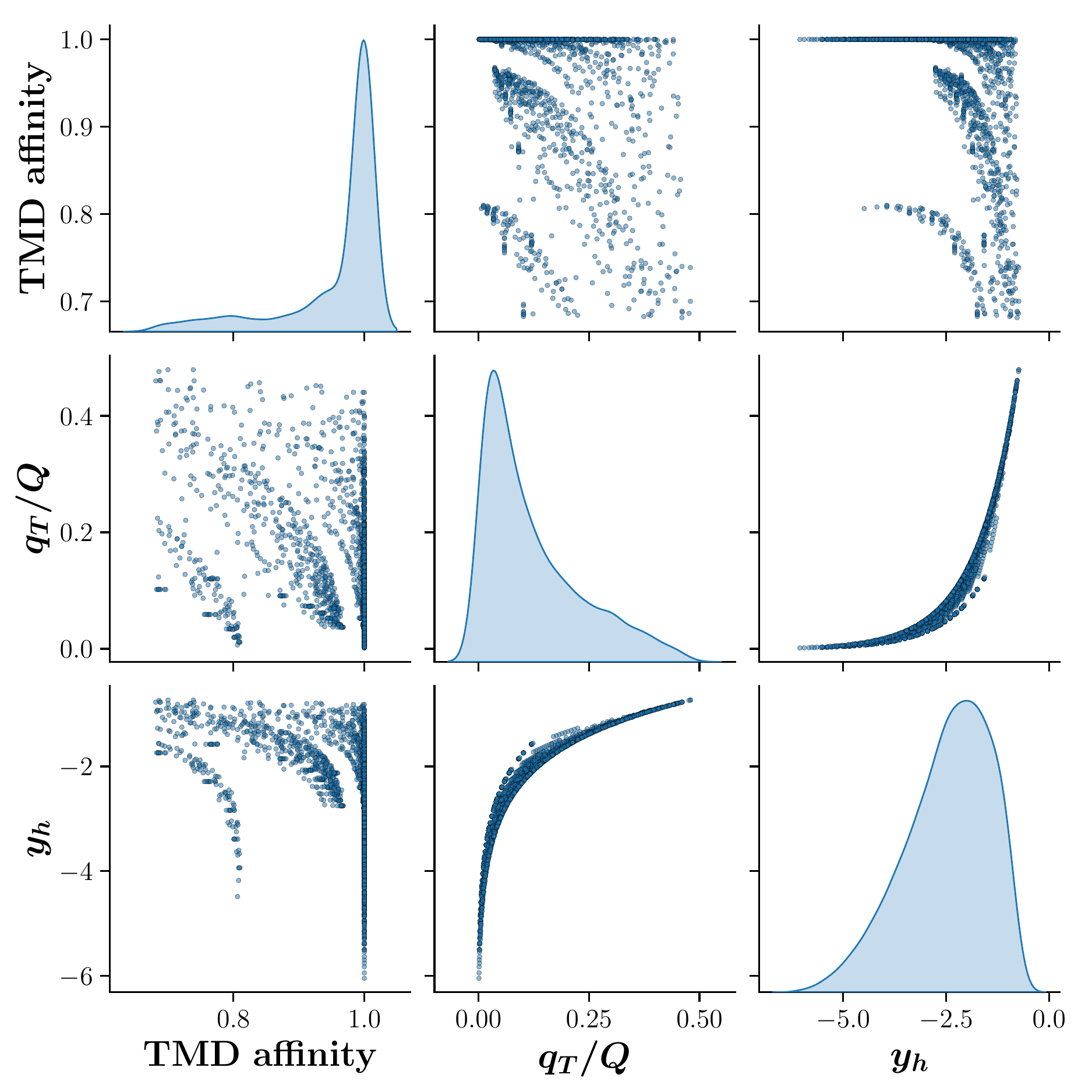}
\caption{Pair-wise correlations among TMD affinity greater than 70\%, $q_T/Q$, and the produced hadron rapidity $y_h$ for EIC kinematics. The histograms on the diagonal help to visualize the relative abundance of points in other pair-wise plots.}
\label{f.EIC_tmd_sb}
\end{figure}

In \fref{EIC_tmd_sb} we show correlation plots of TMD affinity, the ratio $q_T/Q$, and the rapidity~$y_h$. 
The correlation between $y_h$ and $q_T/Q$ appears very strong.
In particular, we observe that the more negative the produced hadron rapidity $y_h$, the lower the values of $q_T/Q$, which is typical of the TMD factorization region. 
Using our settings for the TMD affinity definition, namely $R_{0,1,2} < 0.3$, we obtain a TMD affinity of 68\% or larger for $y_h\in (-6,0)$ and $q_T/Q\lesssim 0.4$.
From our results we find that hadrons in the negative $y_h$ rapidity region are likely to have low values of $q_T/Q$ and large values of TMD affinity.

Both $q_T/Q$ and the rapidity $y_h$ appear to be good proxies for TMD affinity, especially when used in tandem.
In fact, taking into account only a single indicator may considerably limit the accuracy in establishing the region boundaries.
This was also reported in Ref.~\cite{Boglione:2021wov}, where an algorithm based on region indicator ratios involving both $q_T/Q$ and $y_h$ was introduced for the study of the $e^+ e^-\to h X$ process.
Moreover, as the correlation between the rapidity $y_h$ and $q_T/Q$ is very strong, it will be important to obtain information on $y_h$ directly from experimental measurements.

As mentioned above, in this analysis we do not attempt a phenomenological description of the experimental data.
However, it is instructive to compare the affinity data selection to those of existing analyses, such as those presented in Refs.~\cite{Scimemi:2019cmh, Bacchetta:2019sam}. 
If we apply the kinematic cuts as in Eq.~\eqref{eq:cutsVladimirov} to the EIC data, we find 2116 bins survive from the total of 7400.
This subset can be compared to the number of bins that correspond to TMD affinity of 68\% or higher, which is 2325. 
Therefore, only 344 bins do not correspond to cuts from Ref.~\cite{Scimemi:2019cmh}, and 94\% of data selected by cuts from Ref.~\cite{Scimemi:2019cmh} belong to the TMD region with affinity of 68\% or higher.

In the next-to-leading-logarithmic  
precision analysis of SIDIS, Drell-Yan, and $Z$-boson production data in Ref.~\cite{Bacchetta:2017gcc}, the following selection criteria were used:
\begin{align}
Q^2 > 1.4 \,{\rm GeV}^2\, , \quad 
0.2 < z_h < 0.74\, , \nonumber \\
P_{hT} < {\rm min} \big[ 0.2 \, Q,\, 0.7 \, z_h Q \big] + 0.5 \; \rm {GeV}\, .
\label{eq:cutsPavia}
\end{align}
Applying these cuts to the sample of projected EIC data in our analysis, we find 2148 bins (from 7400) are selected, and 1504 of those have TMD affinity of 68\% or higher.
In addition, 838 of these bins do not belong to the bins selected by  Eq.~\eqref{eq:cutsVladimirov} in Ref.~\cite{Scimemi:2019cmh}.

Other selection criteria used in Refs.~\cite{Anselmino:2013lza, Cammarota:2020qcw} for leading order TMD phenomenology~are
\begin{align}
Q^2 > 1.63\,{\rm GeV^2}, \quad
0.2 < z_h < 0.6\, , \quad 
0.2 < P_{hT} < 0.9\; \rm GeV.
\label{eq:cutsJam}
\end{align}
Note that from the point of view of factorization proofs the conditions $q_T \ll Q$ and $P_{hT} \ll Q$ are equivalent.
However, depending on the numerical value for $z_h$, data which satisfy $P_{hT} \ll Q$ may not satisfy $q_T \ll Q$, and may therefore be difficult to describe in a TMD approach.
Applying the cuts (\ref{eq:cutsJam}), we find that 671 bins survive, and 396 of those belong to TMD affinity of 68\% or higher.
Interestingly, only 50\% of the data selected by the cuts from Refs.~\cite{Anselmino:2013lza, Cammarota:2020qcw} overlaps with the data selected by cuts from Refs.~\cite{Scimemi:2019cmh, Bacchetta:2017gcc} in the case of the EIC bins we study.

\begin{figure}[t] 
\centering
\includegraphics[width=1\textwidth]{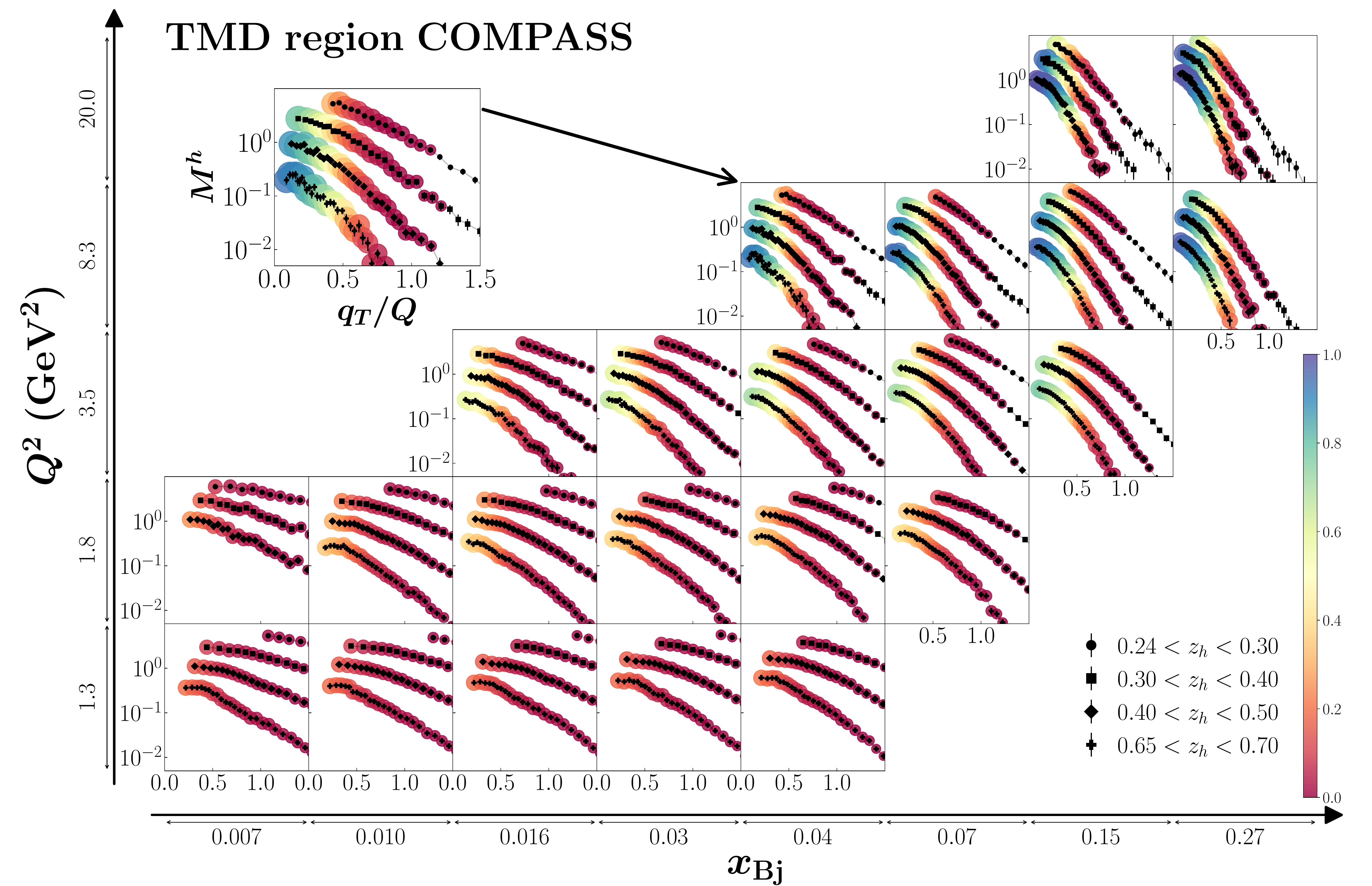}
\caption{TMD region for COMPASS data~\cite{COMPASS:2017mvk} on multiplicities $M^h$ for hadrons $h^+$ (black symbols) produced in muon-deuteron SIDIS, as a function of $q_T/Q$. In each panel there are up to four bins in $z_h$, with the dataset at the top (bottom) corresponding to the lowest (highest) range in $z_h$.}
\label{f.COMPASS_tmd}
\end{figure}

The COMPASS Collaboration performed measurements~\cite{COMPASS:2017mvk} of charged hadrons produced in collisions of 160~GeV longitudinally polarized muons scattered off a deuterium target at typical SIDIS kinematics, for
$Q^2 > 1 \,{\rm GeV}^2$, 
$W^2 \equiv (P+q)^2 > 25\,{\rm GeV^2}$, 
$0.003 < \xb< 0.7$, 
$0.2 < z_h < 1$, and 
$0.1 < y < 0.9$, where 
$y = P\cdot q / P\cdot \ell$. 
The multiplicity in Ref.~\cite{COMPASS:2017mvk} is defined as the ratio of SIDIS to inclusive DIS cross sections,
\begin{align}
\label{Eq:multiplicity-COMPASS}
M^h \equiv 
\frac{\diff^4\sigma_{\rm SIDIS}/\diff{\xb}\, \diff Q^2 \diff z_h\,\diff P_{hT}^2}
     {\diff^2\sigma_{\rm DIS}/\diff\xb\,\diff Q^2} \; .
\end{align} 
In \fref{COMPASS_tmd} we present the kinematic bins 
covered by COMPASS and the data corresponding to the positive hadron multiplicity.
The TMD affinity is represented by the colored circles superimposed over each data point.
In each bin we plot the data for four $z_h$ bins indicated in the legend as a function of $q_T$. 
As in the case of the EIC, higher-$Q$ bins have higher TMD affinity for low values of $P_{hT}$. 
For larger values of $z_h$ and 
$\xb$ one expects higher TMD affinity, as seen in \fref{COMPASS_tmd}.

The COMPASS data were also used in the phenomenological study in Ref.~\cite{Scimemi:2019cmh}. 
For the case of $h^+$ production, 138 bins survive after the cuts defined in Eq.~\eqref{eq:cutsVladimirov}. 
We find 200 bins actually have TMD affinity of 68\% or higher, and 81 of them do not survive after applying Eq.~\eqref{eq:cutsVladimirov}. 
At the same time, 1165 bins are selected by Eq.~\eqref{eq:cutsPavia}, but only 200 of them have TMD affinity of 68\% or higher, while 767 bins are selected by Eq.~\eqref{eq:cutsJam}, but only 106 have TMD affinity of 68\% or higher. 
The conclusion drawn from this is that
additional phenomenological work is needed to delineate the TMD region more precisely.

The HERMES Collaboration measured~\cite{HERMES:2012uyd} the multiplicity of pion and kaon production in the scattering of 27.6 GeV positrons  
from proton and deuteron targets in the SIDIS kinematics 
$Q^2 > 1$~GeV$^2$,
$W^2 > 10$~GeV$^2$,
$0.023 < \xb < 0.4$,
$0.2 <z_h <0.7$, and
$y < 0.85$.
The measured multiplicity in Ref.~\cite{HERMES:2012uyd} was defined as
\begin{align}
\label{Eq:multiplicity-HERMES}
	M_n^h  \equiv 
	\frac{\diff^4\sigma_{\rm SIDIS}/\diff\xb\,\diff Q^2\,\diff z_h\,\diff P_{hT}}
	{\diff^2\sigma_{\rm DIS}/\diff\xb\,\diff Q^2} \; .
\end{align}
In \fref{HERMES_tmd} we show the bins explored by HERMES and the data corresponding to the positive pion multiplicity.
In each bin we plot the data for the $z_h$ values indicated in the legend as a function of $P_{hT}$. 
The TMD affinity is represented by the colored circles superimposed over each data point.
We observe that the affinity is larger for higher $z_h$ and $\xb$ values, and for relatively small $P_{hT}$.
The HERMES data were also used in the phenomenological study of Ref.~\cite{Scimemi:2019cmh}. 
For the case of $\pi^+$ production from a proton target, we find that 34 bins survive after the cuts in Eq.~\eqref{eq:cutsVladimirov}, and 36 bins have TMD affinity of 68\% or higher.

\begin{figure}[t] 
\centering
\includegraphics[width=1.1\textwidth]{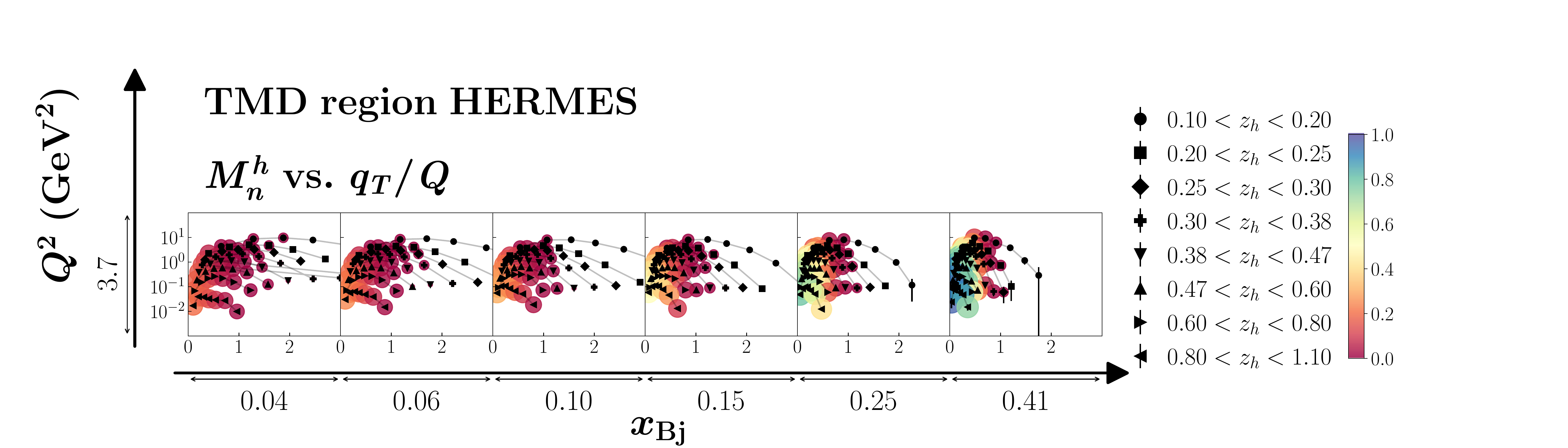}
\caption{TMD region for HERMES multiplicities $M_n^h$~\cite{HERMES:2012uyd} (black points) as a function of $P_{hT}$ for $\pi^+$ produced from a hydrogen target. In each panel there are up to eight $z_h$ bins with the top (bottom) dataset corresponding to the lowest (highest) values of $z_h$.}
\label{f.HERMES_tmd}
\end{figure}
\begin{figure}[t] 
\hspace*{-0.9cm}\includegraphics[width=1.1\textwidth]{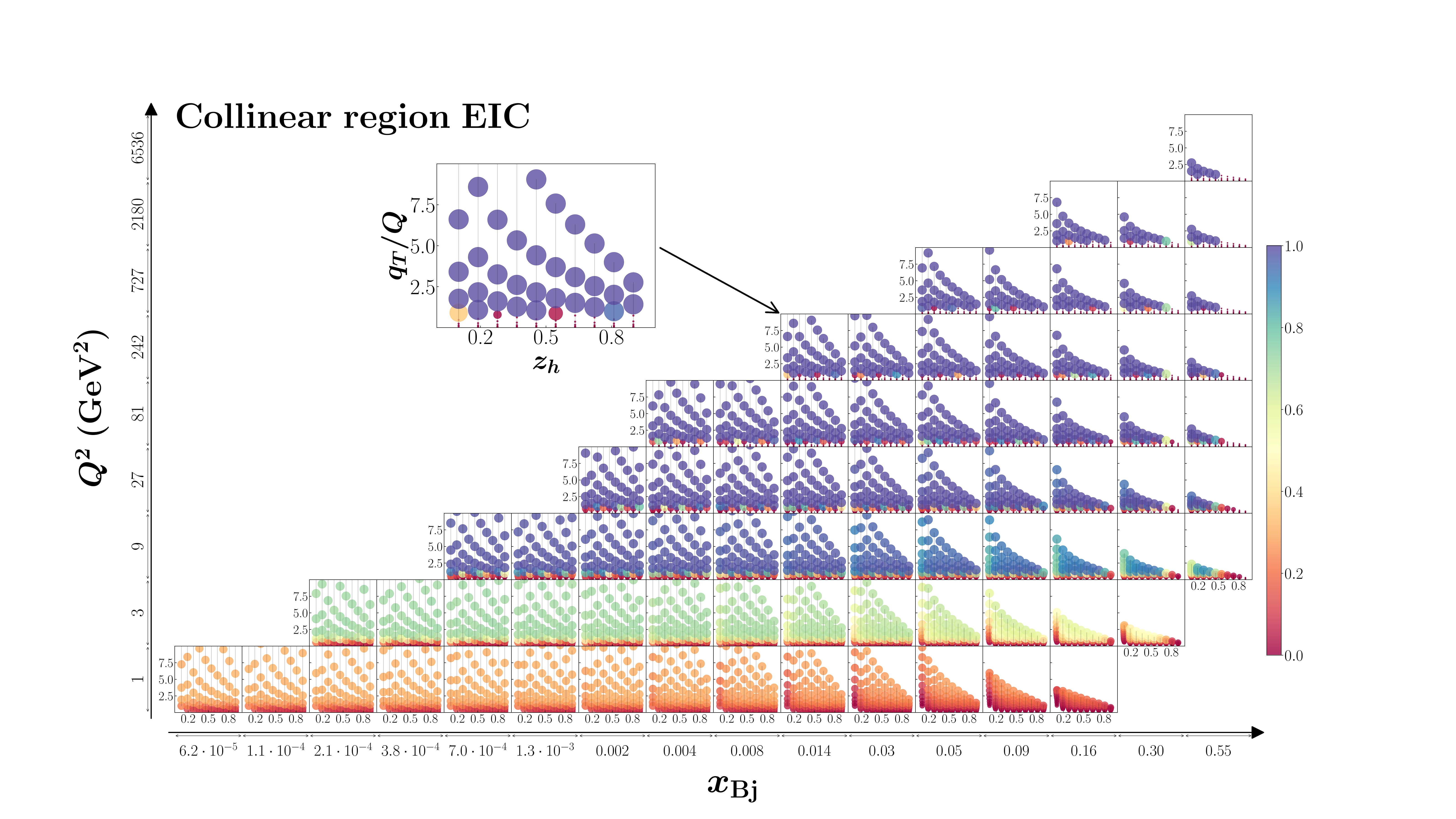}
\caption{Collinear affinity for EIC kinematics. Bin centers are located in the points corresponding to the bin average values of $\xb$ and $Q^2$ (GeV$^2$). In each of these bins, various values of $z_h$ and $q_{T}/Q$ can be measured. In each bin of fixed $z_h$ and $q_{T}/Q$ we plot the affinity as a dot with size proportional to the corresponding affinity value. Affinity is also color coded, according to the scheme on the right of the figure panel: red (and smaller) symbols correspond to low TMD affinity while  dark blue (and larger) symbols to high collinear affinity.}
\label{f.EIC_col}
\end{figure}

\subsection{Collinear region}

\begin{figure}[t] 
\centering
\includegraphics[width=1\textwidth]{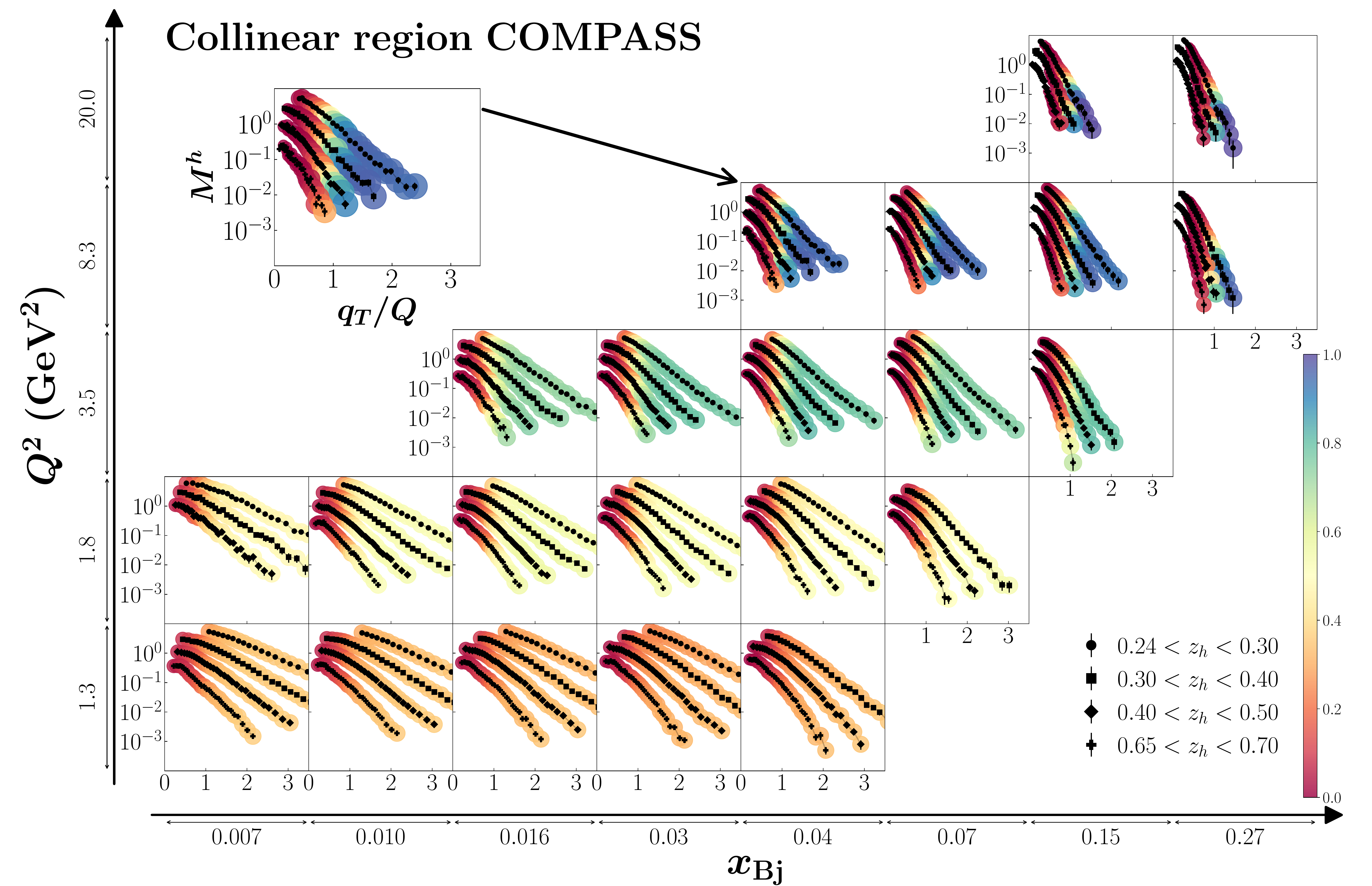}
\caption{Collinear region for COMPASS multiplicities~\cite{COMPASS:2017mvk}. The data (black symbols) on $M^h$ for $h^+$ obtained in muon-deuteron SIDIS are shown as a function of $q_T/Q$. In each panel there are up to four bins in $z_h$, with the dataset at the top (bottom) corresponding to the lowest (highest) range in $z_h$ and vice versa.
}
\label{f.COMPASS_col}
\end{figure}

The collinear current fragmentation region is complementary to the TMD region. 
It covers the region of current fragmentation where hard parton recoil is important, and has negligible sensitivity to the parton intrinsic transverse momentum. 
In terms of region indicators, here the ratio $R_2$ becomes large, while $R_4$ remains small. 
This region has been discussed recently in Refs.~\cite{Wang:2019bvb, Gonzalez-Hernandez:2018ipj}, where a significant tension was found between data and theory at COMPASS and HERMES kinematics, with deviations up to an order of magnitude.
Wang~{\it et al.}~\cite{Wang:2019bvb} showed that such deviations are marginally improved by the inclusion of ${\cal O}(\alpha_s^2)$ corrections. 
Similar observations have been made by Bacchetta {\it et al.}~\cite{Bacchetta:2019tcu} in the context of the analysis of Drell-Yan  cross sections differential in the transverse momentum of the lepton pair.

With the affinity tool in hand, we can now examine the interpretation of existing and future data in the large transverse momentum regime.  
In \fref{EIC_col} we present the affinity results at EIC kinematics, showing the ranges for $q_T/Q < 10$ to focus on the larger transverse momentum regime.
We estimate that 1750 out of the 7400 bins have a collinear affinity of 68\% or higher, while 1170 bins have a collinear affinity of 95\% or higher.
As expected, the collinear affinity is larger for increasing values of $Q$ and $q_T/Q$, while becoming smaller for $q_T/Q < 1$. 
Note that for our chosen values of parton monenta, the collinear affinity values at low $Q$ are not large even when $q_T/Q > 1$, which indicates that in general the $q_T/Q \sim 1$ criterion to estimate the transition from TMD to collinear physics cannot be assumed at low values of $Q$.

\begin{figure}[t] 
\centering
\hspace*{0.5cm}
\includegraphics[width=1.0\textwidth]{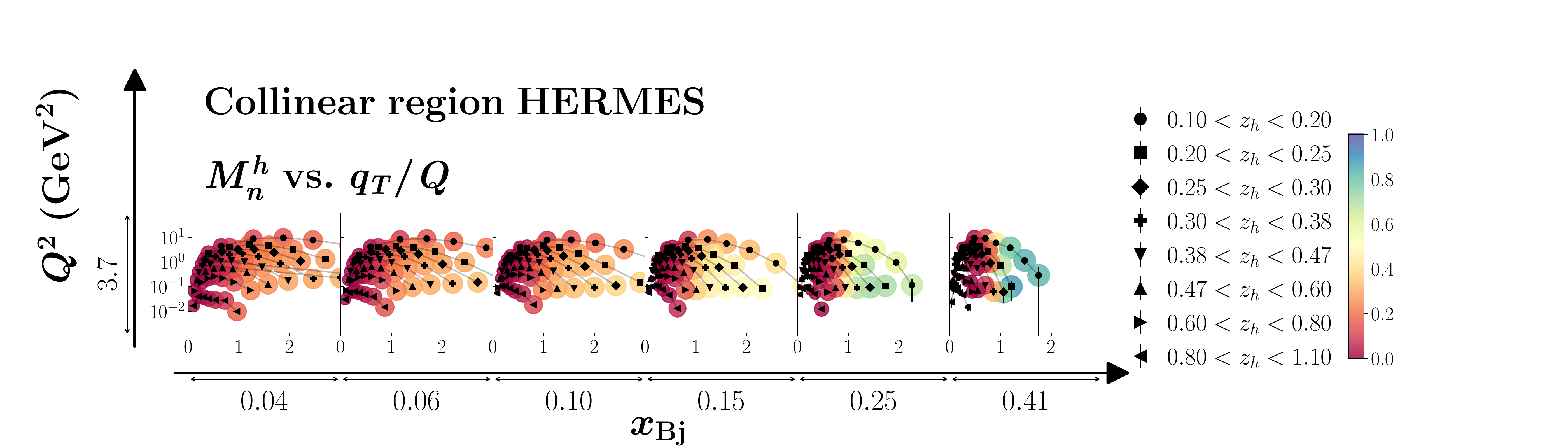}
\caption{Affinity to the collinear region for the HERMES multiplicity data~\cite{HERMES:2012uyd}.}
\label{f.HERMES_col}
\end{figure}
\begin{figure}[t] 
\vspace*{1cm}
\centering
\includegraphics[width=0.49\textwidth]{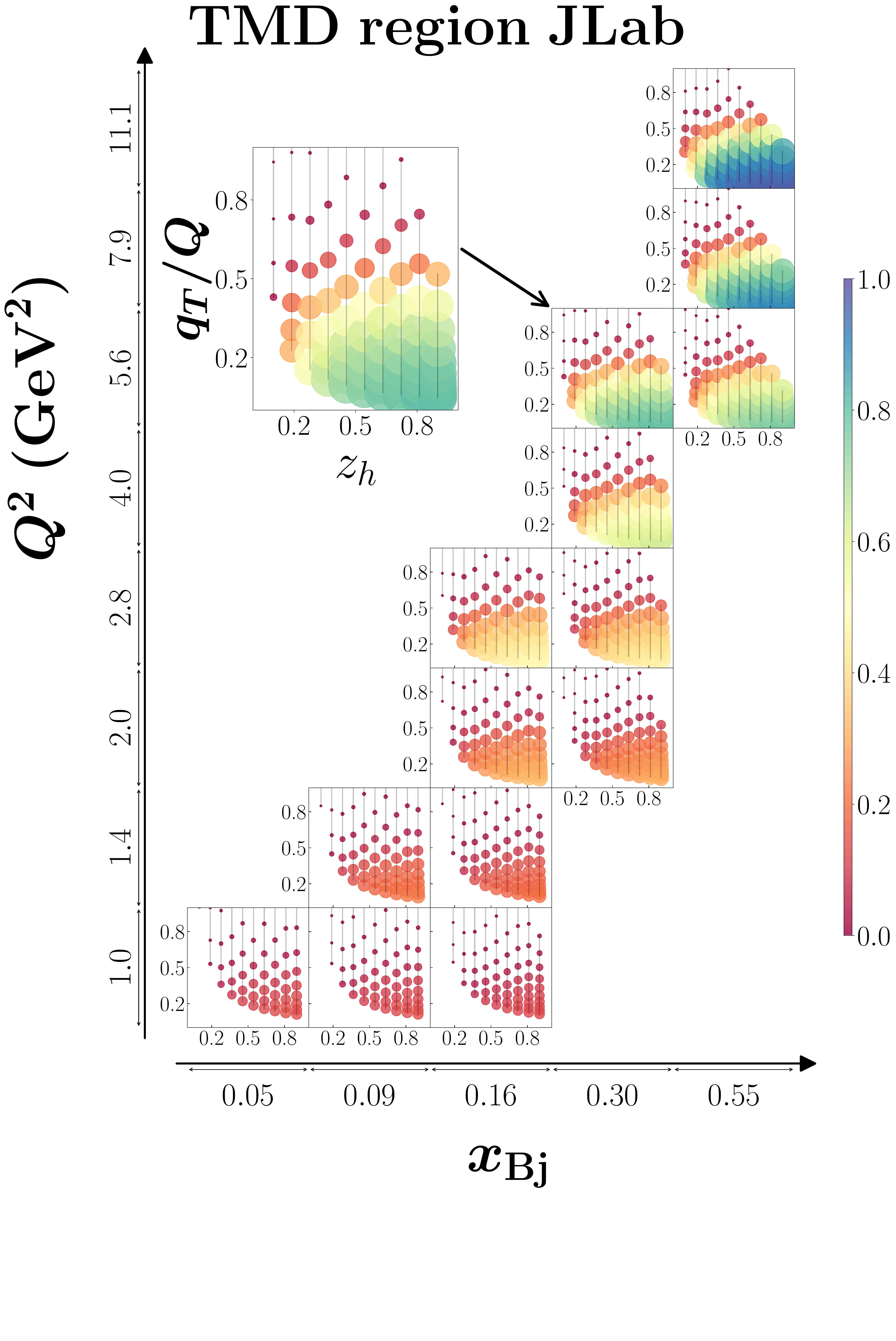}
\includegraphics[width=0.49\textwidth]{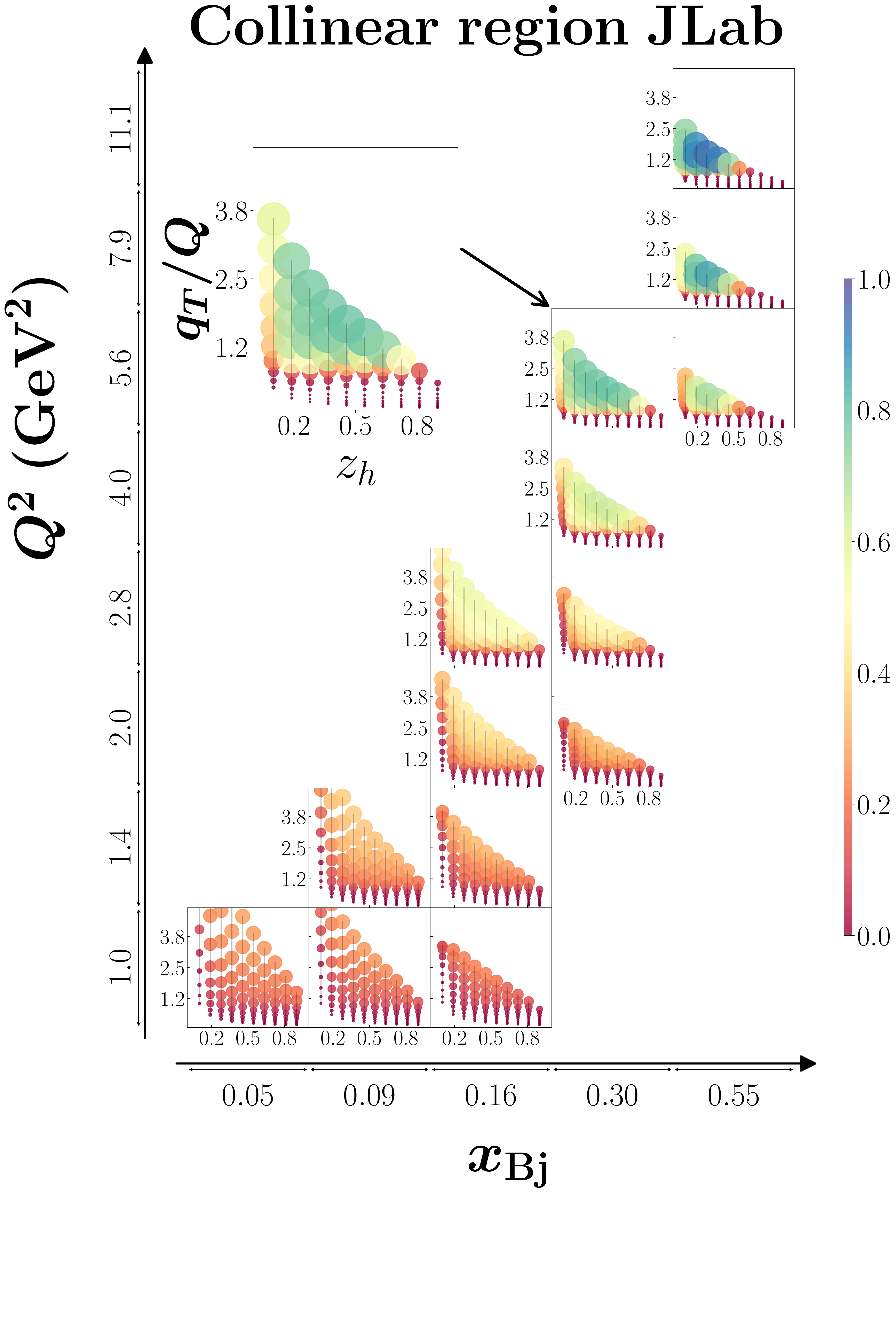}
\vskip -1.5cm
\caption{Affinity to the TMD and collinear regions at Jefferson Lab kinematics.}
\label{f.EIC_jlab}
\end{figure}

The collinear affinity values at COMPASS kinematics are shown in \fref{COMPASS_col}, along with the actual experimental multiplicities. 
Interestingly, the affinity values are in most cases rather small even for $q_T/Q > 1$, which is consistent with the tension between data and theory found in Refs.~\cite{Wang:2019bvb, Gonzalez-Hernandez:2018ipj}.
In contrast, the affinity values become notably larger for high values of $Q$.
However, those regions are close to the edge of the phase space, and it is likely that threshold effects need to be taken into account to achieve a satisfactory description of the cross section in this region~\cite{Kulesza:2003wn}.   
Similar results can be found in Figs.~\ref{f.HERMES_col} and \ref{f.EIC_jlab} at  HERMES and Jefferson Lab kinematics, respectively.
It is evident that here no kinematic region shows a strong affinity to collinear factorization. 
We stress, however, that these observations are based on our specific choices of parton momenta, and in general the results should be viewed only as rough estimates. %

\subsection{TMD-collinear matching region}

\begin{figure}[t] 
\hspace*{-0.9cm}\includegraphics[width=1.1\textwidth]{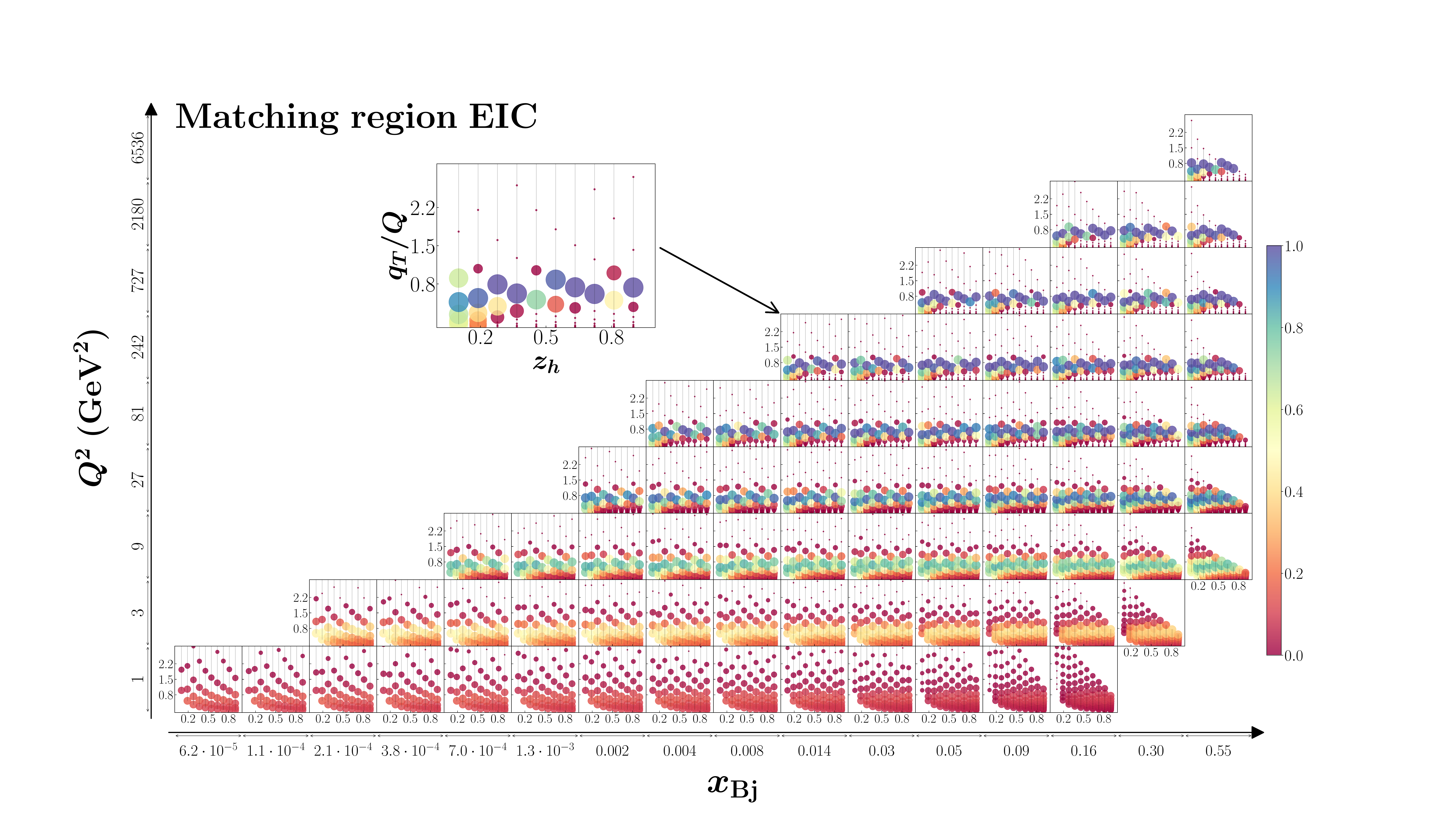}
\caption{Affinity for the TMD-collinear matching region for EIC kinematics. Bin centers are located in the points corresponding to the bin average values of $\xb$ and $Q^2$. In each bin of fixed $z_h$ and $q_T/Q$ we plot the affinity as a dot with size proportional to the corresponding affinity value. Affinity is also color coded, according to the scheme on the right of the figure panel.}
\label{f.EIC_match}
\end{figure}

The TMD-collinear matching region covers a range of $q_T$ values such that $\Lambda_{\rm QCD} \ll q_T \ll Q$, and represents the region where one expects a smooth transition between the TMD and collinear regimes. 
In this intermediate region a description of the data may be possible in either the TMD or collinear schemes. 
Traditionally, in this region one would implement the $W+Y$ construction by CSS~\cite{Collins:1989gx}, which should ensure a smooth cross section over a wide region of $q_T$, with controllable error.
The existence of such a region is one of the important requirements of the CSS formalism.

In fact, matching the SIDIS cross section across the TMD and collinear regions turns out to be rather challenging, especially with regard to lower energy experiments, as discussed in Refs.~\cite{Boglione:2016bph, Collins:2016hqq, Echevarria:2018qyi}, where different matching procedures have been proposed. 
While such a discussion goes beyond the scope of this paper, 
it is important to stress that once a certain definition of this region is chosen, our affinity algorithm can identify and correctly map it, in exactly the same way as for the TMD and collinear regions.

In \fref{EIC_match} we show the TMD-collinear matching region for EIC kinematics, as determined by the affinity tool.
As expected, it correctly covers the range of intermediate values of $q_T$, and turns out to be relevant at rather large values of $Q^2$ corresponding to moderate and large values of $\xb$.

\subsection{Target and central regions}

\begin{figure}[t] 
\hspace*{-0.9cm}\includegraphics[width=1.1\textwidth]{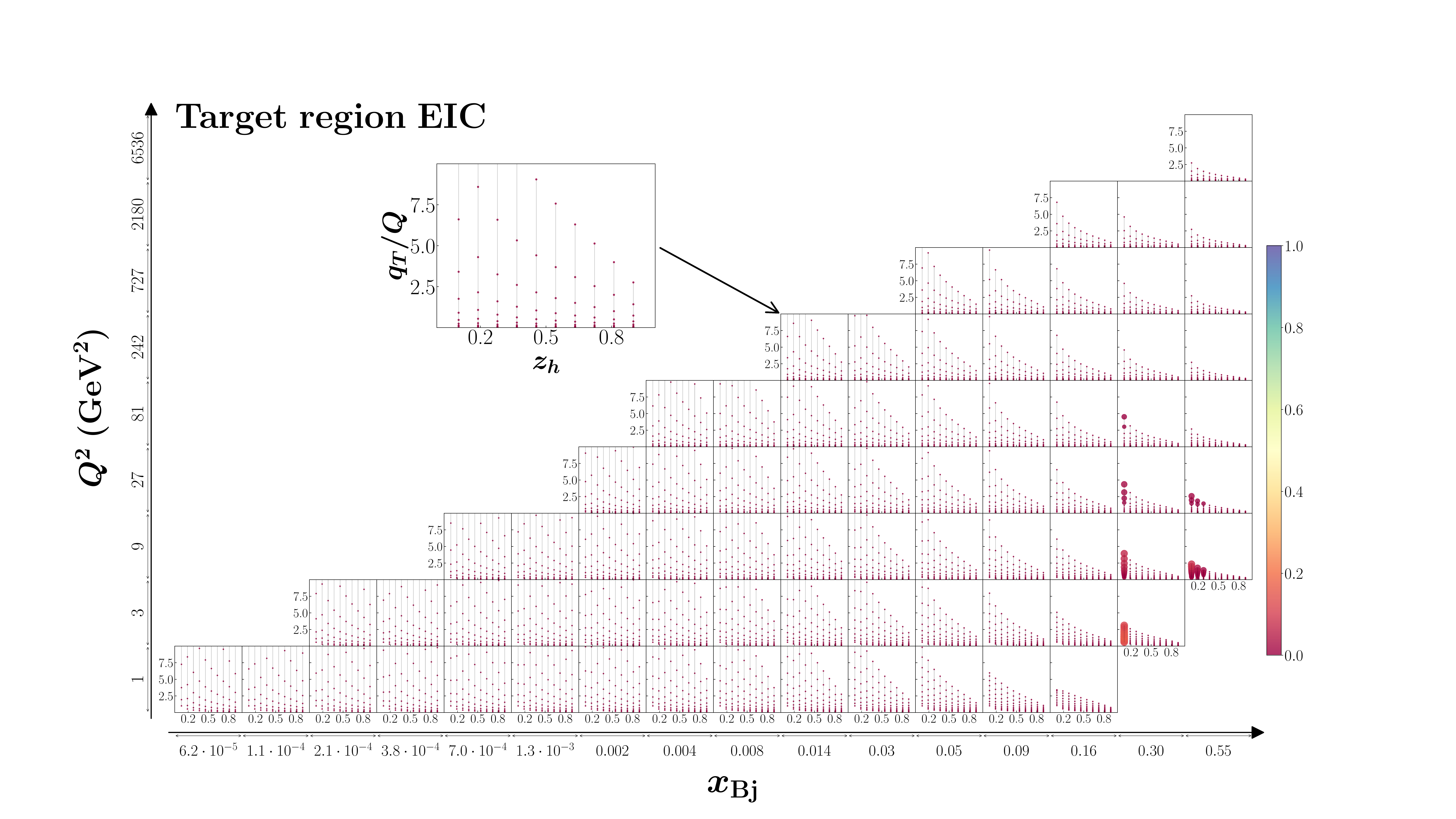}
\caption{Affinity for the target fragmentation region at EIC kinematics.}
\label{f.EIC_target}
\end{figure}
\begin{figure}[t] 
\hspace*{-0.9cm}\includegraphics[width=1.1\textwidth]{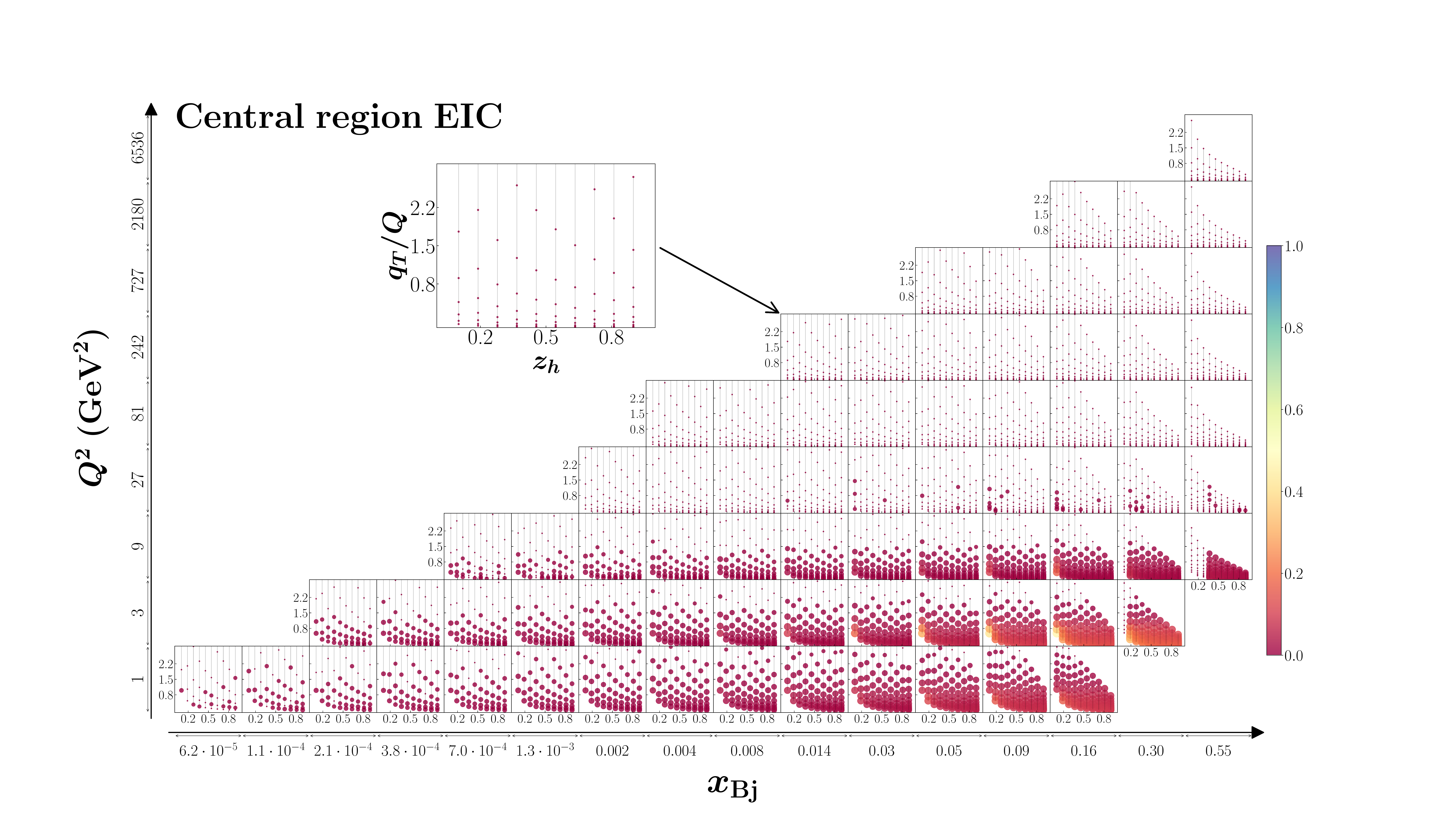}
\caption{Affinity for the central fragmentation region at EIC kinematics.}
\label{f.EIC_soft}
\end{figure}

According to our estimates, at EIC kinematics only a relatively small number of bins is expected to be associated with the target and central fragmentation regions.
Indeed, only 15 bins for the target region and 457 bins for the central region exceed an affinity of 5\%.
The target and central fragmentation regions for the bins of EIC are shown in Figures~\ref{f.EIC_target} and \ref{f.EIC_soft}.

As discussed in Section~\ref{sec:regions}, partons that do not undergo an interaction with the virtual photon hadronize and move predominantly in the direction of the nucleon. 
These target fragmentation hadrons will be found in the region of positive rapidity, close to the beam.
While the experimental measurement of such hadrons is challenging, the study of target fragmentation is important both phenomenologically and theoretically.
These processes are usually described in terms of fracture functions~\cite{Trentadue:1993ka, Grazzini:1997ih, Anselmino:2011ss, Chai:2019ykk}, which are conditional probabilities of producing a hadron  the remnant of a nucleon that carries a fraction $\approx 1-\xb$ of the nucleon's momentum. 
For such hadrons the notion of $z_h$ is not well defined, as in this case $z_h \approx 0$.
Fracture functions are important ingredients in the description of diffractive hadroproduction, and there are attempts to derive factorization formalisms for those processes~\cite{Berera:1995fj}.
Factorized formulas for TMD fracture functions were conjectured in Ref.~\cite{Anselmino:2011ss}, and the evolution equations along with a more detailed study of factorization was proposed in Ref.~\cite{Chai:2019ykk}.
Ref.~\cite{Anselmino:2011ss} also derives correlations that can be studied experimentally. 
All this body of work will be important
for the planning of EIC detectors~\cite{AbdulKhalek:2021gbh}.

Our estimates of affinity to the target fragmentation region for EIC kinematics are shown in \fref{EIC_target}.
One can see that the target region is characterized by relatively large values of $\xb$ and small values of $z_h$.
Obviously, a more detailed study of the target fragmentation region is needed in order to fully realize the potential of the EIC.
Such a study may include a detailed Monte Carlo model of the hadrons produced in the target fragmentation region together with {\tt Geant4} detector simulations.

The last region we will discuss in this section is the central fragmentation region. 
It is known that the region in rapidity between the struck quark and the nucleon remnants will be filled with radiation that is needed to neutralize the color and make the production of colorless hadrons from colored partons possible. 
In the event generators in Ref.~\cite{Sjostrand:2016bif} one employs a Lund string model that describes fragmentation as the fracturing of a flux tube created between by the colored quark and the remnant of the nucleon (see Refs.~\cite{Andersson:1983ia, Andersson:1997xwk} and references therein).
As a result, the rapidity between the produced hadron and the remnants of the nucleon is filled with hadrons. 
It is of course interesting to reconcile fragmentation models, especially ones that include spin, such as the Lund string model~\cite{Kerbizi:2018qpp} or the Nambu--Jona-Lasinio model of fragmentation~\cite{Ito:2009zc, Matevosyan:2016fwi}, with results of QCD.
A recent attempt to use Feynman-graph structures was presented in Ref.~\cite{Collins:2018teg}.
In addition, the central region is relevant for factorization, as soft radiation plays an important role for proofs of factorization, including TMD factorization~\cite{Collins:2011zzd, Collins:2016ztc}.

Even though the central region is incorporated in Monte Carlo generators used in experimental analyses, this region has not yet been explored in detail empirically, and future experimental studies, such as of rapidity distributions of hadrons, will be needed. We estimate that for the central region we have a majority of events in the low-$Q^2$ and low-$P_{hT}$ region, as indicated in \fref{EIC_soft}. 
These hadrons are present in the region of central rapidities, $y_h \sim 0$, and as we explored in \fref{exp_dist}, the contribution to this region of rapidity comes from most of the fragmentation regions investigated in this paper. 
We expect that in the future more detailed experimental, theoretical, and phenomenological studies will allow more precise delineations of this region~\cite{Collins:2016ztc, Collins:2018teg}.

\section{Interactive affinity tool}
\label{sec:interactive}

The calculation of affinity is numerically demanding and time consuming.
In order to facilitate the computations we use machine learning techniques to train a neural network model for fast evaluation of affinity.

\begin{figure}[t] 
\hspace*{-0.5cm}
\begin{center}
\includegraphics[width=0.85\textwidth]{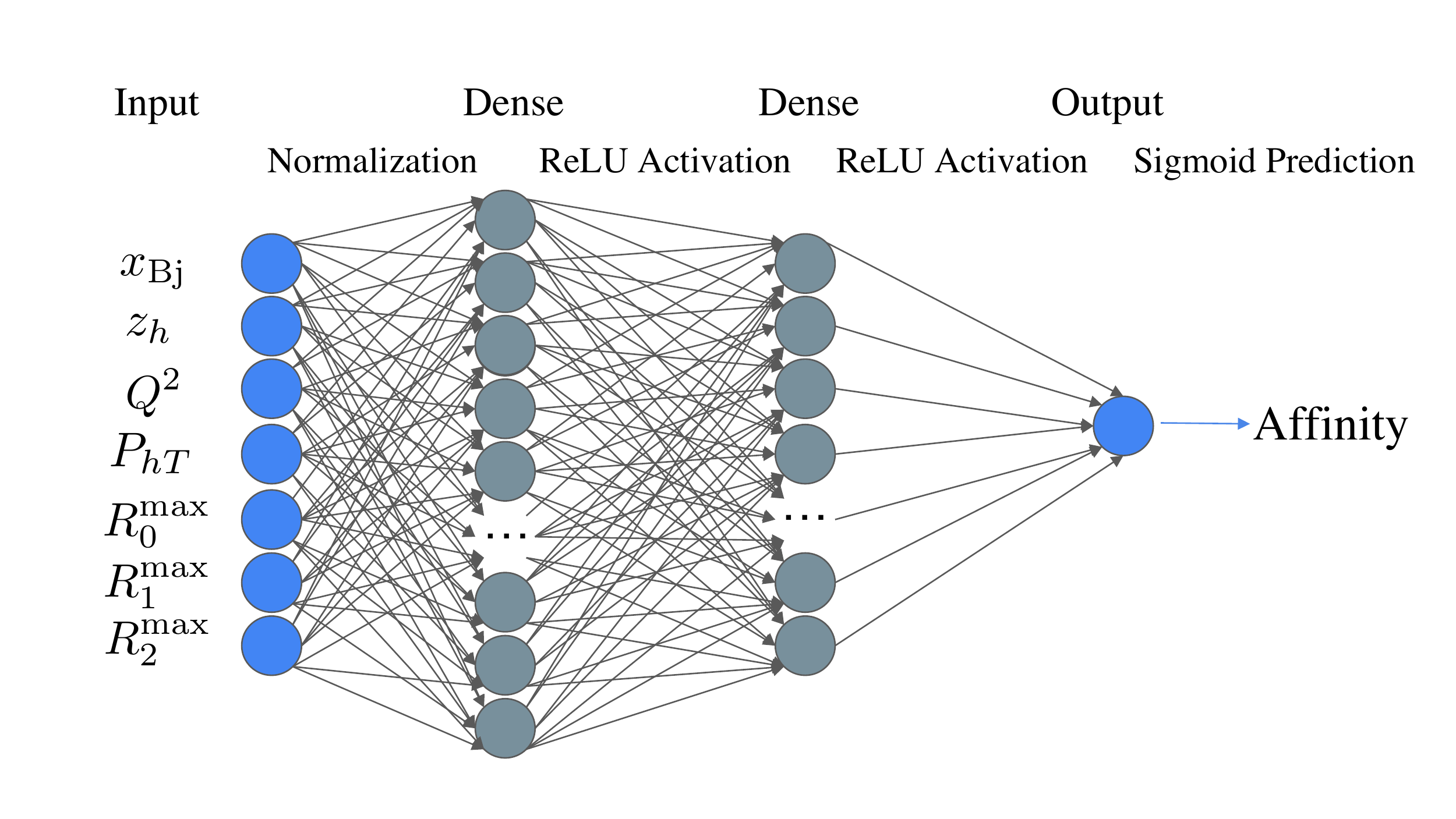}
\hspace*{-1cm}
\end{center}
\caption{Schematic illustration of the architecture of the neural network devised for calculating affinity. Each arrow represents a trainable parameter that each neural network must learn through training. Input, hidden layers, and output are represented by the (stack of) blobs. The normalization layer and ReLU activation functions are necessary to mitigate the vanishing gradient problem. (See text for a more detailed explanation.)}
\label{f.net}
\end{figure}

Firstly, we generate the affinity data that will be used in training the neural networks by varying the maximum values of three ratios, $R_0$, $R_1$, and $R_2$, and keeping other nonperturbative parameters fixed, as discussed in the paper.
The affinity is defined with respect to the values $R_0^{\rm max}$, $R_1^{\rm max}$, and $R_2^{\rm max}$ varying each in the range $(0.05, 1.25)$.
We generate 
$\approx 15000$ configurations\footnote{This number gave very good predictive power of the models.} of those parameters, and for each configuration of these values we produce the corresponding set of affinities for each of the 7400  EIC experimental bins.

We use the {\tt TensorFlow}~\cite{tensorflow2015-whitepaper} framework to create and train four neural networks that predict the TMD, collinear, target, and central affinity regions with seven input values of $\xb$, $z_h$, $P_{hT}$, $Q^2$, $R_0^{\rm max}$, $R_1^{\rm max}$, and $R_2^{\rm max}$. 
Choosing the best neural network is not always straightforward or deterministic, as there are many hyperparameters to be adjusted, such as the number and the width of hidden layers, as well as the minimization algorithms to be used.
A good hyperparameter combination can significantly improve the performance of the network, and we employ the hyper-band {\tt KerasTuner}~\cite{omalley2019kerastuner} for tuning the hyperparameters for our networks.
The repository that contains the resulting neural network models can be accessed at a {\tt GitHub} repository \cite{affinity-repo}.

We train four separate neural nets to predict the affinity value to the TMD, collinear, target, and central regions.
{\tt KerasTuner} hyperparameter search results in each net consisting of four layers: the input layer, two hidden layers, and the output layer.
A pictorial representation of the architecture of this network is presented in \fref{net}, where arrows correspond to the internal parameters that each neural net must learn through training, and the input, output and hidden layers are represented by the (stack of) blobs. 
Training is an iterative process of updating weights by providing the model input data accompanied by affinity values.

The first hidden layer has 576 neurons for the TMD, 960 for the collinear, 896 for the target, and 576 for the central regions. 
The second hidden layer has 160 neurons for the TMD, 544 for the collinear, 256 for the target, and 736 for the central regions. 
For neuron activation we use the rectified linear activation (ReLU) function $f(x) = \max (0,x)$, where $x$  
is the input of an activation layer, or the weighted sum of each node in that layer. 
Here, the weights are trainable parameters and the sum runs over the output of the last layer.
The normalization layer and ReLU activation functions are necessary to mitigate the vanishing gradient problem; in fact, when back propagation fails, the weights cannot be updated.

\begin{figure}[t] 
\centering
\includegraphics[width=0.45\textwidth]{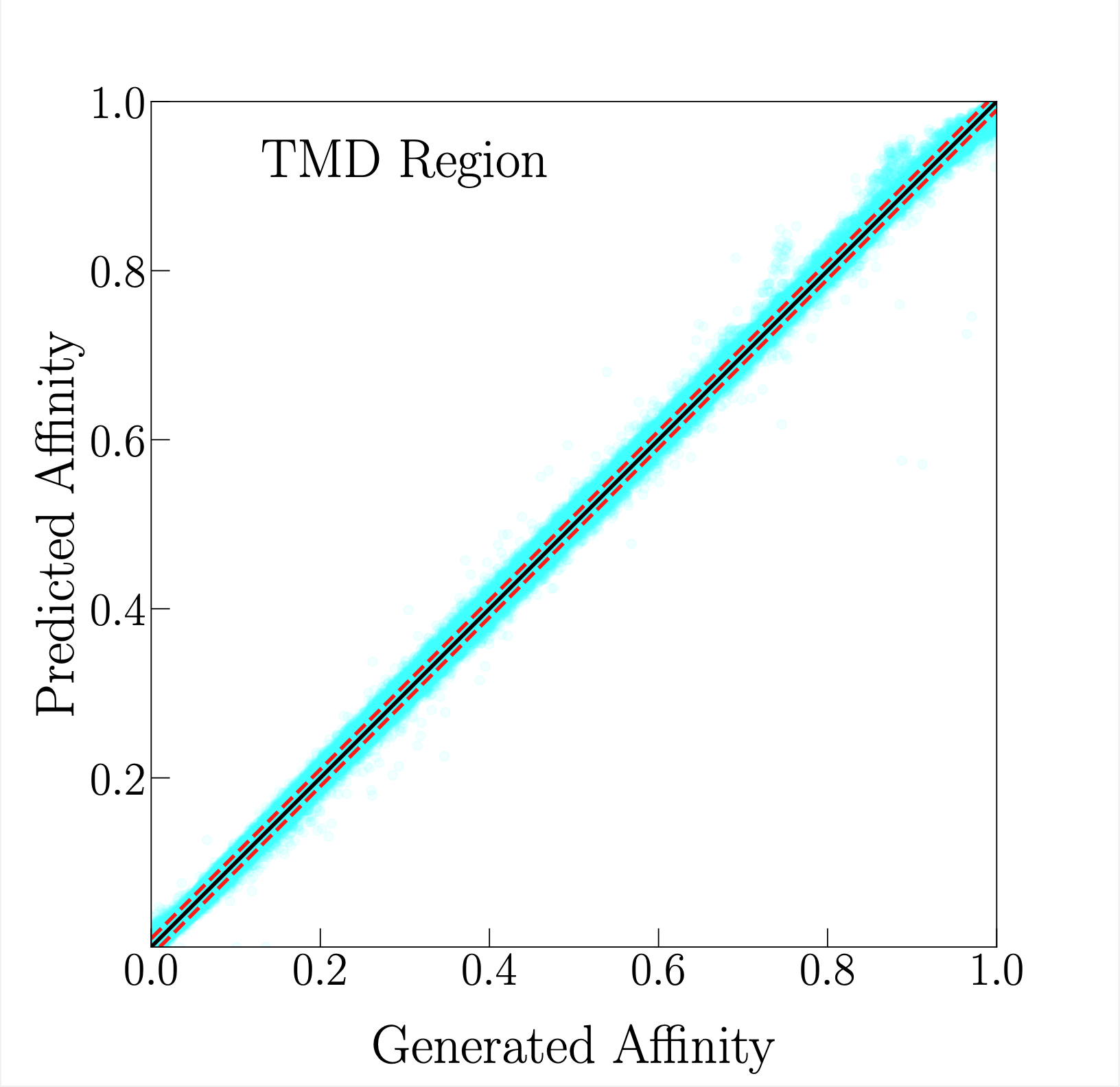} \hspace{1.cm}
\includegraphics[width=0.45\textwidth]{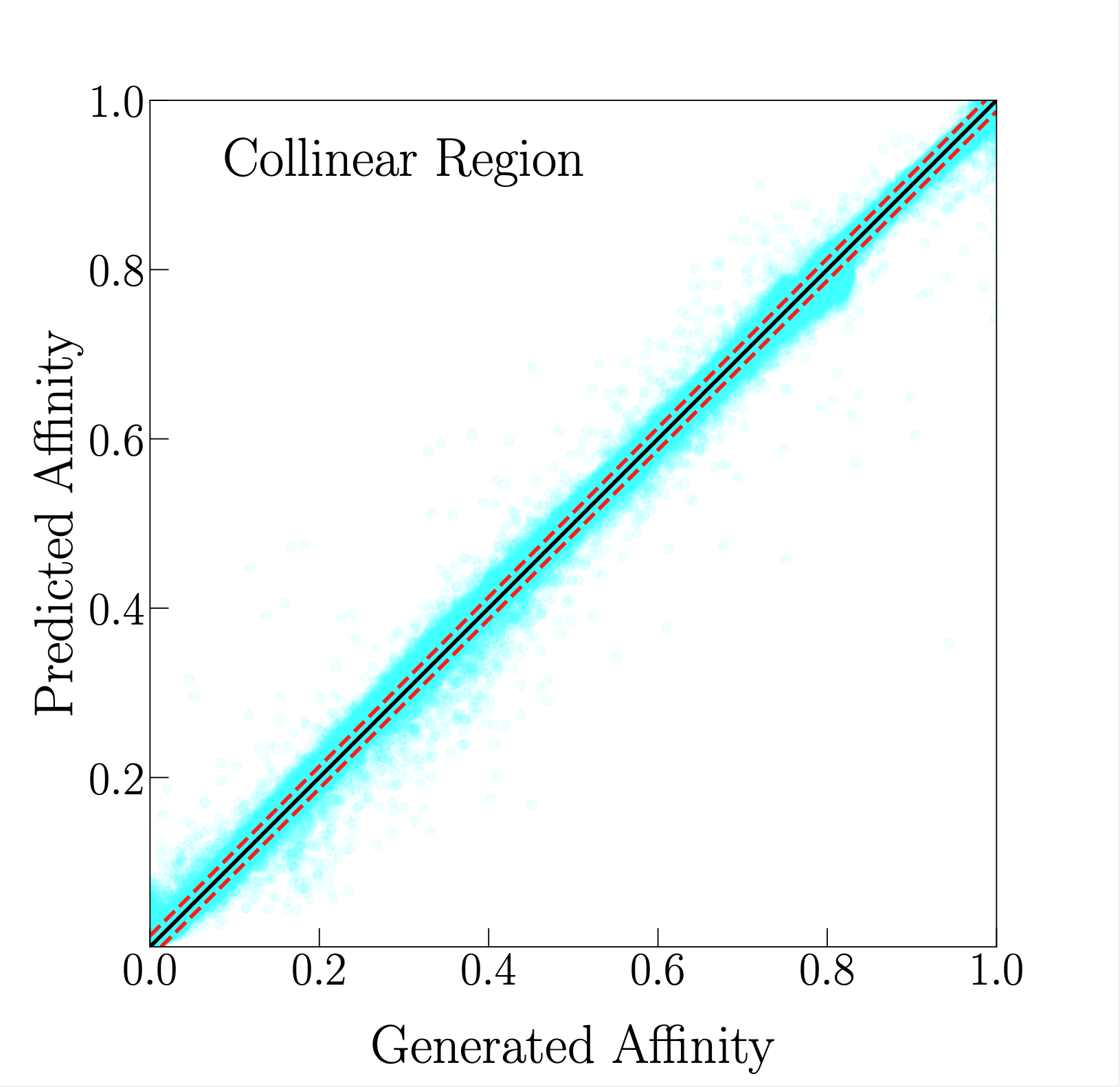}\\ 
\hspace{-0.1cm}
\includegraphics[width=0.435\textwidth]{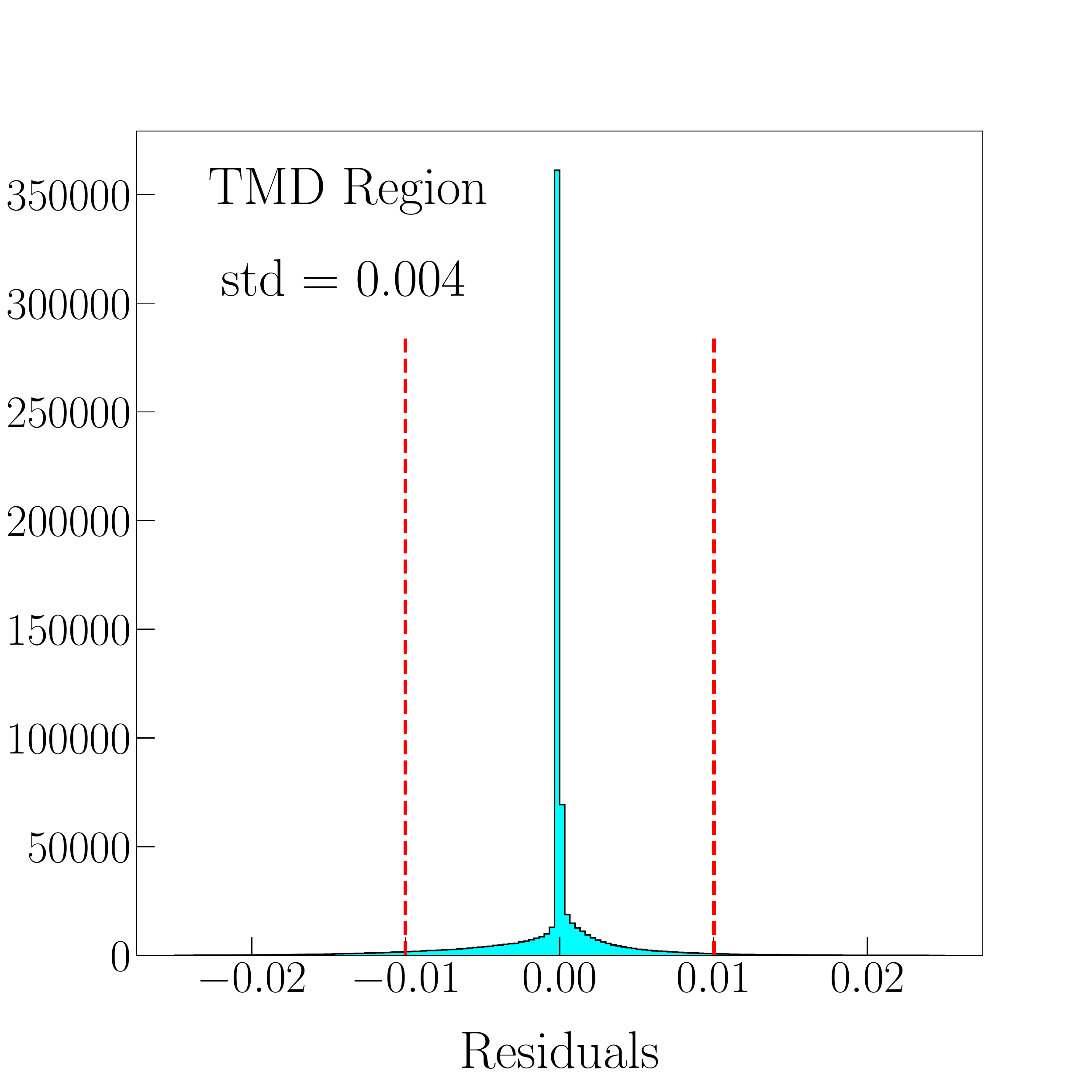} \hspace{1.3cm}
\includegraphics[width=0.435\textwidth]{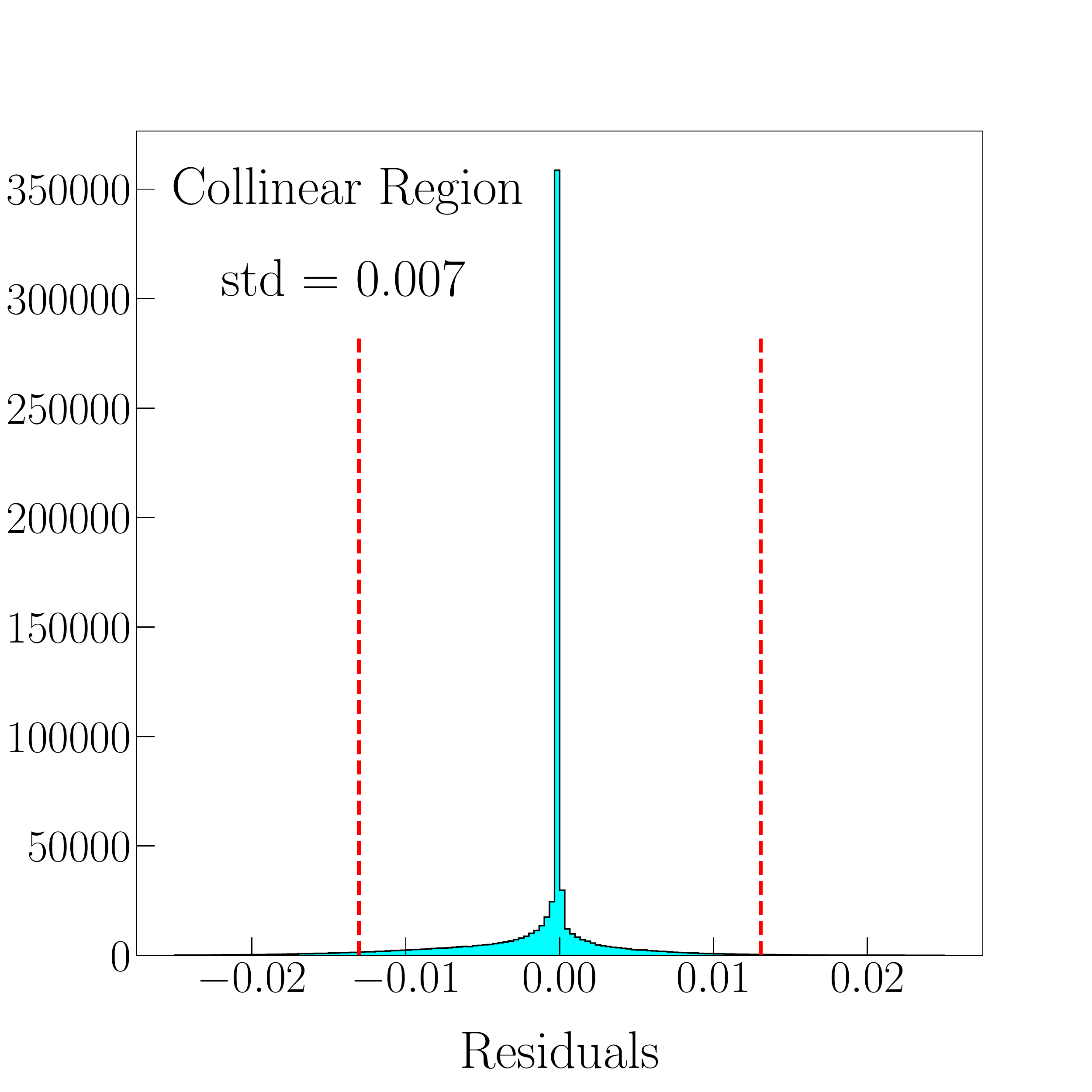}
\caption{Upper panels: Scatter plots of predicted affinity versus the affinity generated with test data. The left (right) panel corresponds to the TMD (collinear) affinity. In each panel the black solid central line represents a perfect prediction, corresponding to zero residuals, while the points located between the two parallel red dashed lines contain 95\% of all residuals.
Lower panels: Histograms of residuals for the TMD (left) and collinear (right) regions. The red dashed lines showing the intervals that contain 95\% of all residuals, with the standard deviation ``std'' indicated. Note that the actual range of the residual values is wider than that shown in the histograms, but the outliers in the plots are omitted for better visual clarity.}
\label{f.prediction-test}
\end{figure}

To obtain predictions of affinities bounded on the interval from 0 to 1, we choose the activation function of the last layer (output) as the so-called sigmoid function, which maps any real value to the range $(0, 1)$.
This is a cumulative distribution function of a logistic distribution, and transforms the weighted sum of the second dense layer output into a probability or confidence of prediction.
Other output functions were also considered in the tuning process.

We used 80\% of the data for training and 20\% for validation of the network learning, which provides a set of input data to test the network at the end of each training iteration.
This procedure allows one to avoid over-fitting and to check the accuracy of the predictions.
We use the validation predictions to calculate the mean squared error (MSE), for which back propagation minimizes.
Minimization of the validation loss, performed with the Adam minimizer, optimizes the weight parameters. 
The model architecture that provides the smallest validation MSE (loss function) was considered to be the best.
Each model training lasted between 50 and 500 epochs (iterations over the training data).
After the training, the resulting neural networks were saved in order to be available for use in further {\tt TensorFlow} applications.
For each hyper search, the networks with the five smallest validation MSE were summarized.
When the best network was unclear, an independent test set was used to inform the architecture chosen to be further trained and implemented in the affinity tool.

Finally, 100 datasets for each of the 7400 experimental bins and corresponding affinities were independently generated for testing, which allowed for visual and measurable comparisons of networks.
The predicted and generated affinity values shown in \fref{prediction-test}  provide a visualization of the error distribution, with a center-line representing perfect predictions and a cloud of points depicting residuals (differences between the values of generated and predicted affinity) for the TMD and collinear network models.
The scatter plots in the upper panels show two parallel red lines, equally spaced about the black line of zero residuals, which contains 95\% of all residuals.  
The cloud of points shows higher variation in the TMD network error for larger values of affinity compared to smaller values, while the variation of error for collinear appears to be more uniform over the range of affinity values. 
Histograms of residuals are shown in the lower panels of~\fref{prediction-test}. 
Around 11\% of all predictions have zero residuals, resulting in the histograms being sharply peaked.
The standard deviations, 0.004 for TMD affinity and 0.007 for collinear affinity, confirm a very good accuracy of predictions of the trained neural network models.

The validated and tested affinity tool has been made publicly available.
To minimize the need for installation of additional software for the affinity tool users, we utilize a {\tt Google~Colab} notebook that allows users to launch the tool from a web browser without additional installation.
The notebook clones the corresponding {\tt GitHub} repository that contains the neural networks and all scripts that produce the interactive plots.
The user can choose values for $R_0^{\rm max}$, $R_1^{\rm max}$, and $R_2^{\rm max}$ and generate plots corresponding to the TMD, target, central, and collinear regions at the EIC using neural network models trained for these regions.
The interactive notebook with user instructions can be found in Ref.~\cite{affinity-tool}.

\section{Conclusions}
\label{sec:conclusions}

SIDIS measurements offer a tremendous opportunity to learn about the partonic structure of nucleons.
For a correct phenomenological interpretation of the information they encode, it is vital to develop tools that allow experimental data to be connected to the corresponding theoretical framework. 
Factorization theorems only apply under specific kinematic conditions, essentially dictated by power counting.
It is therefore very important to be able to identify as precisely as possible the sensitivity of each data subset to those kinematic requirements.

In this paper we have implemented the region indicators, $\{R_i$\}, first introduced in Refs.~\cite{Boglione:2016bph, Boglione:2019nwk}, to quantify our confidence in the proximity of SIDIS observables to a particular physical mechanism.
For this purpose we have devised a new tool, ``affinity'', to facilitate the separation of phase space regions where different factorization formalisms apply.
We quantify affinity by combining information from the Monte Carlo generation of partonic configurations and the resulting ratios $\{R_i$\} into a single estimate of proximity to a particular hadron production region, which ranges from 0$\%$ to 100$\%$. 
The affinity to the TMD current fragmentation region is estimated for HERMES and COMPASS datasets for unpolarized multiplicities, and for Jefferson Lab and EIC kinematics.  
We also quantify the proximity of the current fragmentation region for large transverse momenta described by a collinear QCD treatment, and the transition region from the TMD to collinear factorization descriptions~\cite{Collins:2016hqq}.
The central and target regions are also addressed, but these require further phenomenological investigation and dedicated theoretical studies.

Our affinity tool shows that a large portion of experimental bins can be associated with either TMD or collinear physics, for all considered experiments, and especially for the future EIC.
Lower energy experiments such as those at Jefferson Lab, however, show a non-negligible admixture of central and target fragmentation events. 
The affinity tool can be applied in phenomenological analyses to select kinematic bins that are sensitive to the kinematic region of interest. 
It can also be used to guide the development of new SIDIS experiments, and to incorporate the region estimator into experimental analyses.
For this reason we also provide a publicly available interactive tool that allows the study of affinity according to any choice for the separation of the kinematic regions.
This tool is based on a neural network model trained with machine learning techniques which allows a fast evaluation of the affinity. 
The architecture of the net consists of four layers: input, output, and two hidden intermediate layers. 
The affinity interactive tool is available as a {\tt Google~Colab} notebook~\cite{affinity-tool}, which can easily be accessed and run from any browser without the need of additional software. \\

\acknowledgments

We would like to thank Ted Rogers for helpful discussions and collaboration in the early stages of this research.
This work has been supported by the National Science Foundation under Grants No.~PHY-2011763 (D.P.), No.~PHY-2012002 (A.P., S.D., Z.S.), the U.S. Department of Energy, under contracts No.~DE-FG02-07ER41460 (L.G.) and No.~DE-AC05-06OR23177 (M.D., A.P., N.S., W.M.) under which Jefferson Science Associates, LLC, manages and operates Jefferson Lab, and within the framework of the TMD Topical Collaboration. A.P. would like to thank Temple University for hospitality and support during his sabbatical leave.
The work of N.S. was supported by the DOE, Office of Science, Office of Nuclear Physics in the Early Career Program.
M.B. acknowledges funding from the European Union’s Horizon 2020 
research and innovation programme under grant agreement STRONG – 2020 - No 824093”.

\clearpage

\providecommand{\href}[2]{#2}\begingroup\raggedright\endgroup

\end{document}